\documentclass[prb,showpacs,twocolumn,aps,superscriptaddress,floatfix]{revtex4-1}
\usepackage{amsmath}
\usepackage{amssymb}
\usepackage{bm}
\usepackage{graphicx}
\usepackage{color}
\usepackage[svgnames]{xcolor}
\usepackage{soul}
\usepackage{hyperref}
\usepackage{verbatim}% Long comments

\DeclareMathOperator{\Real}{Re}
\DeclareMathOperator{\Div}{div}

\begin{document}

\title{High frequency permeability of the composite with ferromagnetic spherical shells}

\author{A.O. Sboychakov}
\affiliation{Institute for Theoretical and Applied Electrodynamics, Russian Academy of Sciences, 125412 Moscow, Russia}

\begin{abstract}
The paper studies high-frequency permeability of the composite materials consisting of hollow ferromagnetic particles embedded into the non-magnetic media. We model the ferromagnetic particles in composite by spherical shells: the thickness of the ferromagnetic region $d$ compared to the  particles' diameter $D$ can vary in a wide range, from $d\ll D$ to $d\sim D$. We assume that the magnetization distribution in such a particle is non-uniform, but forms a vortex-like structure: the magnetization is twisted in some plane outside two vortex cores placed at the poles of the particle. We consider two types of magnetic anisotropy, which help to stabilize such a magnetic configuration: the easy-plane magnetic anisotropy and the ``circular'' uniaxial magnetic anisotropy with easy axis rotated together with the magnetization direction outside the vortex core. The high-frequency permeability of such a composite material has been studied in the limit of non-interacting particles. We study the dependence of the permeability on the ratio $d/D$. We showed that for $d\sim D$ the composite's permeability behaves in similar manner for both types of magnetic anisotropy. It was shown also that for the second type of magnetic anisotropy and in the limit $d/D\ll1$ the frequency dependence of the particle's susceptibility is quite similar to that for the thin film. At the same time, the magnetization oscillations in the ac field are non-homogeneous for both types of anisotropy.
\end{abstract}

% 75.78.-n	Magnetization dynamics
% 75.75.Jn	Dynamics of magnetic nanoparticles
% 75.75.-c	Magnetic properties of nanostructures

\date{\today}
\maketitle

\section{Introduction}

Magnetic composites are interesting both from a fundamental point of view and for a large number of technical applications, for many of which it is important to know the system's response to an external alternating magnetic field. The internal structure of composite materials is extremely diverse. In this paper, we will be interested in materials consisting of non-contacting ferromagnetic particles with spherical or near spherical shape, separated by a non-magnetic dielectric matrix. Due to the peculiarities of the manufacturing technology, it often turns out that ferromagnetic particles contain a hollow region in their centers~\cite{LI2015131,C7CP03292G,Sui2018,coatings10100995,s22083086,magnetism3020008}. The characteristic sizes of such particles $D$ can vary in a wide range. For example, in Ref.~\onlinecite{LI2015131}, the particles in the composites synthesized had sizes of about $500$\,nm with a characteristic thickness of the ferromagnetic region $d\sim80$\,nm. Similar~\cite{coatings10100995,magnetism3020008}, or slightly larger~\cite{s22083086} ($D\sim1$\,$\mu$m) particles were obtained in materials studied in Refs.~\onlinecite{coatings10100995,s22083086,magnetism3020008}. Composites with particles of larger~\cite{C7CP03292G} ($D\sim 30$--$100$\,$\mu$m) and smaller~\cite{Sui2018} ($D\sim 100$\,nm) sizes have been also synthesized. The ratio of the thickness of the ferromagnetic layer $d$ to the particle's size $D$ can also vary in a wide range. Thus, according to the Table~2 of Ref.~\onlinecite{Sui2018}, the ratio of $d/D$ for different samples varied from about $0.1$ to about $0.3$. As for the larger particles, studied in the Ref.~\onlinecite{C7CP03292G}, the $d/D$ ratio could be about $0.05$.

In this paper, we focus on the particle sizes $D\sim0.5$--$1$\,$\mu$m, characteristic of composites synthesized, in particular, in Refs.~\onlinecite{coatings10100995,s22083086,magnetism3020008}. Such particle sizes are not too large, which points in favor of their homogeneous structure. At the same time, for such particle sizes the condition $l_{ex}/D\ll1$ is satisfied with good accuracy, where $l_{ex}$ is the exchange length of the ferromagnet, which is of the order of several nanometers~\cite{LandauVIIIen}. We model the ferromagnetic particles in composite by spherical shells with outer radius $R_2$, and the inner radius $R_1$, while $d=R_2-R_1$ is the thickness of the ferromagnetic region. For further consideration, it will  be convenient for us to introduce the quantity $R_0=(R_2+R_1)/2$ -- the radius of the central layer. Thus, for the particle sizes under consideration, we have $l_{ex}/R_0\sim10^{-2}\ll1$.

\begin{figure}[t]
\centering
\includegraphics[width=0.7\columnwidth]{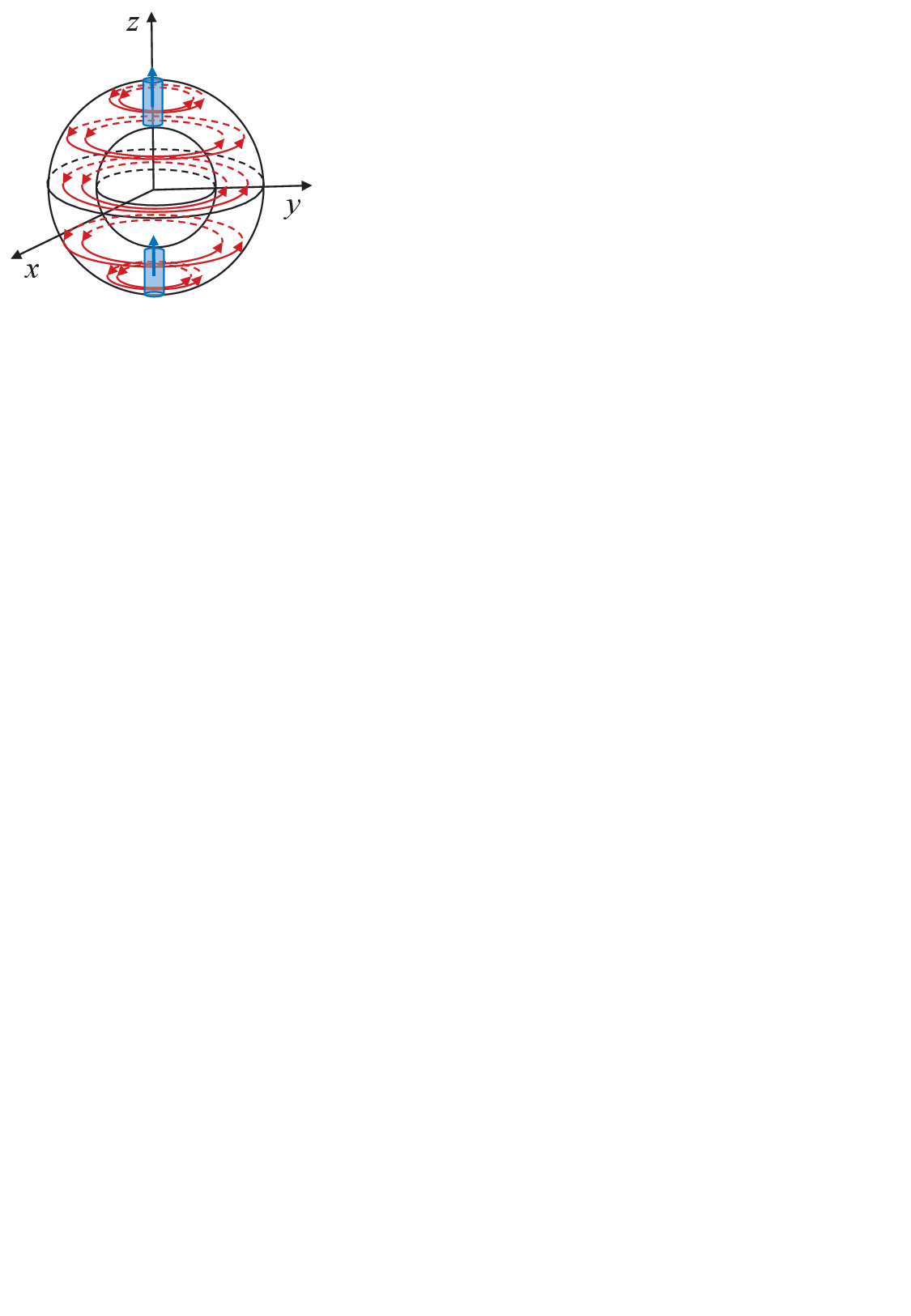}
\caption{\label{FigConfig} The vortex magnetic configuration of the spherical shell. Outside the vortex cores, located at poles of the particle, the magnetization vectors lie in the $xy$ plane and form a system of circles, as shown schematically in the figure.}
\end{figure}

In this paper we assume that the distribution of the magnetization in a particle is non-homogeneous, but forms a vortex-like structure: with the exception of the area near the poles, the magnetization vectors lie in the same plane ($xy$ plane) and are directed parallel to the surface of the particle. Near the poles, there are two domains with magnetization vectors directed parallel to the $z$ axis of the particle. The typical domain sizes are $r_0\times r_0\times d$, where $r_0\sim l_{ex}$. Schematically, this structure is shown in the Fig.~\ref{FigConfig}. For the spherical shells, such a configuration was studied, e.g., in Refs.~\onlinecite{VortexConfiguration,VortexConfiguration2,ModernElectrodynamicsEn2023,VortexConfigurationExp}. For cylindrical particles, it was considered in Ref.~\onlinecite{Usov1993}. In Ref~\onlinecite{VortexConfigurationExp} it has been shown experimentally using the electron holography technique, that the static magnetic configuration of the hollow ferromagnetic particle indeed has the vortex-like structure. Frequency dependence of the permeability for vortex magnetic configuration has been studied theoretically in Ref.~\onlinecite{VortexConfiguration}. However, the particle sizes, studied their ($R_0\sim100$\,nm), are about order of magnitude smaller than that studied in this paper. In our case, as we will show below, the main contribution, determining the frequency dependence of the permeability, is coming from the magnitostatic energy, but not from the exchange energy.

Since the magnetization in a particle is spatially inhomogeneous, the oscillations of the magnetization vectors under the action of an alternating magnetic field will also be inhomogeneous. We take this fact into account when determining the susceptibility of both the particle and the composite as a whole. The plan of this work is as follows. In Section~\ref{LLEq}, we will derive the linearized Landau-Lifshitz equations for the magnetic configuration described above for two types of magnetic anisotropy proposed. In section~\ref{Permeability}, the formula for the magnetic permeability of the composite will be obtained. The results of the numerical calculation of the permeability, as well as the properties of the magnetization oscillations at different frequencies and the directions of the ac field will be analyzed. In section~\ref{PermeabilityLimit}, analytical formulas for the magnetic permeability of the composite in the limit of $d/R_0\to0$ and $l_{ex}/R_0\to0$ will be derived for both types of magnetic anisotropy. In section~\ref{Stability}, the particle's self-oscillations are considered. The stability analysis of the considered configuration is performed. Section~\ref{Discussion} is devoted to the discussion of the results obtained. Some details of the calculations are provided in the Appendix.

\section{Landau--Lifshitz equation for the ferromagnetic spherical shell}\label{LLEq}

We start with the Landau-Lifshitz equation, which has the form~\cite{LandauVIIIen}
\begin{equation}\label{LLeq}
\frac{\partial\mathbf{M}}{\partial t}=\gamma\mathbf{M}\times\frac{\delta U}{\delta\mathbf{M}}+
\frac{\alpha}{M_s}\mathbf{M}\times\frac{\partial\mathbf{M}}{\partial t}-\gamma\mathbf{M}\times\mathbf{h}\,,
\end{equation}
where $\mathbf{M}$ is the magnetization of the particle, $M_s$ is the saturation magnetization, $\gamma$ is the gyromagnetic ratio, $\alpha$ is the decay parameter, $U$ is the total magnetic energy of the particle, and $\mathbf{h}$ is the external alternating magnetic field. The magnetic energy of the particle $U$ includes three terms, $U=U_m+U_a+U_{ex}$, where $U_m$ is the magnetostatic energy, $U_a$ is the energy of the magnetic anisotropy, and $U_{ex}$ is the energy of the exchange interaction. Let us consider first the energy of the magnetic anisotropy. We consider two types of magnetic anisotropy. For both of these types we assume that the energy of the magnetic anisotropy is quadratic on the magnetization vector, so we can write~\cite{LandauVIIIen} [here and further, the summation over the double indices (e.g., $\alpha=x,\,y,\,z$) is implied]
\begin{equation}\label{Ua}
U_{a}=-\frac{K}{2}\int\limits_{V}\!d^3\mathbf{r}\,M^{\alpha}N_{\alpha\beta}M^{\beta}\,,
\end{equation}
where $K$ is the dimensionless anisotropy constant, and $N_{\alpha\beta}$ is the magnetic anisotropy tensor, which, in general, depends on the coordinates. The fist type (type-I) of the magnetic anisotropy we study is just an ease-plane magnetic anisotropy. For this case we have $K<0$, and the anisotropy tensor can be written in the matrix form as (the $xy$ plane is assumed to be the easy plane)
\begin{equation}\label{NphiI}
N^{\alpha}_{\beta}(\mathbf{r})=\left(\begin{array}{ccc}
0&0&0\\
0&0&0\\
0&0&1
\end{array}\right).
\end{equation}
Another type (type-II) of the magnetic anisotropy we consider is the special type of the uniaxial magnetic anisotropy: we assume that at each point the anisotropy axis lies in the $xy$ plane and is directed parallel to the tangent to the particle's boundary, i.e. in the equilibrium configuration shown in Fig.~\ref{FigConfig}, the direction of the magnetization (everywhere except the vortex core) and the anisotropy axis coincide. For this configuration we have $K>0$, and the anisotropy tensor can be written as
\begin{equation}\label{Nphi}
N^{\alpha}_{\beta}(\mathbf{r})=\left(\begin{array}{ccc}
\sin^2\varphi&-\sin\varphi\cos\varphi&0\\
-\sin\varphi\cos\varphi&\cos^2\varphi&0\\
0&0&0
\end{array}\right),
\end{equation}
where $\varphi$ is the azimuthal angle of the spherical coordinate system. Not that the type-II magnetic anisotropy is not solely theoretical construction. For example, some microwires have similar magnetic anisotropy outside their cores~\cite{phan2008giant,zhukov2015advances}. Below we will assume that the case $K<0$ ($K>0$) corresponds to the type-I (type-II) magnetic anisotropy.

The magnetostatic energy has the form~\cite{LandauVIIIen}
\begin{equation}\label{Um}
U_m=-\frac12\int\limits_{V}\!d^3\mathbf{r}\,\mathbf{MH}^\text{in}[\mathbf{M}]\,,\;\;
\mathbf{H}^\text{in}[\mathbf{M}]=-\frac{\partial\Psi[\mathbf{M}]}{\partial\mathbf{r}}\,,
\end{equation}
where $\Psi[\mathbf{M}]$ is the potential of the magnetic field created by the magnetic moments of the particle. It is equal to
\begin{eqnarray}
\Psi[\mathbf{M}]&=&-\int\limits_{V}\!d^3\mathbf{r}'\,\frac{\Div'\mathbf{M}(\mathbf{r}')}{|\mathbf{r}-\mathbf{r}'|}+
\int\limits_{S}\!d\mathbf{S}'\,\frac{\mathbf{M}(\mathbf{r}')}{|\mathbf{r}-\mathbf{r}'|}\nonumber\\
&=&\int\limits_{V}\!d^3\mathbf{r}'\,\mathbf{M}(\mathbf{r}')\frac{\partial}{\partial\mathbf{r}'}\frac{1}{|\mathbf{r}-\mathbf{r}'|}\,.\label{Psi}
\end{eqnarray}
Finally, the exchange energy is~\cite{LandauVIIIen}
\begin{equation}\label{Uex}
U_{ex}=\frac{l_{ex}^2}{2}\int\limits_{V}\!d^3\mathbf{r}\,
\frac{\partial M^{\alpha}}{\partial x^{\beta}}\frac{\partial M_{\alpha}}{\partial x_{\beta}}\,.
\end{equation}
In all formulas above, the three-dimensional integrals are taken over the volume of the ferromagnetic region of the particle $V$, and the surface integral is taken over its inner and outer surfaces.

Since the particle has a spherical shape, it will be convenient for us to switch from the Cartesian coordinate system $x^{\alpha}=\{x,\,y,\,z\}$ to the spherical one $x^{a}=\{r,\,\theta,\,\varphi\}$. In addition, we will also work with vectors in the spherical coordinate system. The latter are related to the vectors in the Cartesian coordinate system by the expressions~\cite{LandauIIen}
\begin{equation}\label{trans}
A^{\alpha}=\frac{\partial x^{\alpha}}{\partial x^{b}}A^{b}\,,\;\;
A^{a}=\frac{\partial x^{a}}{\partial x^{\beta}}A^{\beta}\,.
\end{equation}
We also introduce the metric tensor
\begin{equation}\label{gab}
g_{ab}=\text{diag}(1,\,r^2,\,r^2\sin^2\theta)\,.
\end{equation}
In the spherical coordinate system, the type-I magnetic anisotropy tensor takes the form
\begin{equation}\label{NphiSphericalI}
N^{a}_{b}(\mathbf{r})=\left(\begin{array}{ccc}
\cos^2\theta&-r\sin\theta\cos\theta&0\\
-\sin\theta\cos\theta/r&\sin^2\theta&0\\
0&0&0
\end{array}\right).
\end{equation}
For the type-II magnetic anisotropy tensor we have
\begin{equation}\label{NphiSpherical}
N^{a}_{b}(\mathbf{r})=\left(\begin{array}{ccc}
0&0&0\\
0&0&0\\
0&0&1
\end{array}\right).
\end{equation}
The potential of the magnetic field can be rewritten as
\begin{equation}\label{PsiSpherical}
\Psi[\mathbf{M}]=\int\limits_{R_1}^{R_2}r'^2dr'\!\!\int\!\!d\Omega'\,
M^{a}(\mathbf{r}')\frac{\partial}{\partial x'^{a}}\frac{1}{|\mathbf{r}-\mathbf{r}'|}\,,
\end{equation}
where $d\Omega'=\sin\theta'd\theta'd\varphi'$. The exchange energy will take the form
\begin{eqnarray}
&&U_{ex}=\frac{l_{ex}^2}{2}\int\limits_{R_1}^{R_2}\!\!r^2dr\!\!\int\!\!d\Omega\,M^{a}_{;b}M_{a}^{;b}=\label{UexSpherical}\\
&&\frac{l_{ex}^2}{2}\int\!\!d\Omega\,\left[r^2M^{a}M_{a}^{;r}\right]\Big|_{r=R_1}^{r=R_2}-
\frac{l_{ex}^2}{2}\int\limits_{R_1}^{R_2}\!\!r^2dr\!\!\int\!\!d\Omega\,M^{a}M_{a\phantom{b};b}^{;b}\,,\nonumber
\end{eqnarray}
where, in the latter equation, we clearly distinguished the surface and the volume contributions to the exchange energy. In Eq.~\eqref{UexSpherical}, the symbol `;' stands for the covariant derivative~\cite{LandauIIen}. Finally, in the spherical coordinate system, the Landau-Lifshitz equations have the form
\begin{equation}\label{LLeqSpherical}
\frac{\partial M^a}{\partial t}=\sqrt{g}g^{ab}\epsilon_{bcd}M^{c}\left[\gamma g^{de}\frac{\delta U}{\delta M^{e}}+
\frac{\alpha}{M_s}\frac{\partial M^{d}}{\partial t}- h^{d}\right],
\end{equation}
where $\sqrt{g}=r^2\sin\theta$ and $\epsilon_{bcd}$ is the Levi-Civita symbol ($\epsilon_{r\theta\varphi}=1$).

Our aim is to find a linear response of the system to an alternating magnetic field. To do this, we linearize Eq~\eqref{LLeqSpherical} by representing the magnetization as $M^a=M^a_0+m^a(t)$, where $M^a_0$ is the equilibrium magnetization of the particle. Following Ref.~\onlinecite{ModernElectrodynamicsEn2023}, we will assume that outside the vortex core, the equilibrium magnetization of the particle is
\begin{equation}\label{M0}
M_0^{\alpha}=M_s\left(\begin{array}{c}
-\sin\varphi\\ \cos\varphi\\0
\end{array}\right),\;\;
M_0^a=\frac{M_s}{r\sin\theta}\left(\begin{array}{c}
0\\0\\1
\end{array}\right),
\end{equation}
that is, only the $\varphi$ component of $M_0^{a}$ is nonzero. Next, in Ref.~\onlinecite{ModernElectrodynamicsEn2023} it was also shown that the longitudinal dimensions $r_0$ of the domains located at the poles of the particle are small ($r_0\sim l_{ex}\ll R_0$). Therefore, we will neglect the presence of these domains wherever it is possible. Thus, it is easy to see that ignoring the vortex core, we have $\Psi[\mathbf{M}_0]=0$. The energy of the type-I (type-II) magnetic anisotropy in equilibrium is $U_a=0$ ($U_a=-K M_s^2V/2$). As for the exchange energy, it logarithmically diverges at the poles of the particle, i.e. for $\theta\to 0$ or $\pi$, if we extend the expression~\eqref{M0} to the entire definition area. Therefore, when calculating the exchange energy, we will integrate the diverging term over $\theta$ within the range $l_{ex}/R_0<\theta<\pi-l_{ex}/R_0$, assuming that the contribution to $U_{ex}$ from the vortex core is small. Taking all aforementioned into account, it can be shown that the following equality is satisfied with a good accuracy:
\begin{equation}\label{LL0}
\epsilon_{abc}M^{b}g^{cd}\left.\frac{\delta U[\mathbf{M}]}{\delta M^{d}}\right|_{\mathbf{M}=\mathbf{M}_0}\!\!\!\!\!\!\!=0\,,
\end{equation}
that is, configuration~\eqref{M0} does correspond to some extremum of energy. In Section~\ref{Stability}, we will show that this extremum is a minimum (generally speaking, it is not necessarily global). In addition to Eq.~\eqref{LL0}, the boundary conditions to the Landau-Lifshitz equations, following from the expression for the surface term in the exchange energy~\cite{LandauVIIIen}, are satisfied:
\begin{equation}\label{LL0bound}
\left.\epsilon_{abc}M^{b}M^{c;r}\right|_{\mathbf{M}=\mathbf{M}_0\atop r=R_{1,2}}=0\,.
\end{equation}
Note that Eq.~\eqref{LL0} is valid for both types of magnetic anisotropy considered.

Expanding Eq.~\eqref{LLeqSpherical} around $M^a_0$ to the first order in $m^a(t)$, we can obtain linearized Landau-Lifshitz equations. However, it will be more convenient for us to work not with the vector $m^a$, whose components have different dimensions, but with an object defined as (no summation on $a$)
\begin{eqnarray}
\tilde{m}^a&=&\left(\begin{array}{c}
m^r\\rm^{\theta}\\rm^{\varphi}
\end{array}\right),\;\;\tilde{m}^a=\varkappa^am^a\,,\;\;\varkappa^a=\left(\begin{array}{c}
1\\r\\r
\end{array}\right),\nonumber\\
\tilde{m}_a&=&\frac{1}{\varkappa^a}m_a\,.\label{tildem}
\end{eqnarray}
Objects of the type $\tilde{m}^a$ are not vectors (let's call them ``pseudo-vectors'') in the sense that they are not transformed according to Eq.~\eqref{trans}, but all components of $\tilde{m}^a$ have the same dimension, coinciding with that in the Cartesian coordinate system. Similarly, covariant and contrvariant ``pseudotensors'' of the second rank can be introduced. Substituting formulas~\eqref{Um},~\eqref{NphiSphericalI}\,--\,\eqref{UexSpherical}, \eqref{M0}, and~\eqref{tildem} into Eq.~\eqref{LLeqSpherical} and linearizing it by $\tilde{m}^a$, we obtain the following equations for the type-II magnetic anisotropy
\begin{widetext}
\begin{eqnarray}
&&-\frac{1}{\omega_s}\frac{\partial\tilde{m}^{\theta}}{\partial t}+\frac{\alpha}{\omega_s}\frac{\partial\tilde{m}^{r}}{\partial t}
+K\tilde{m}^r-\frac{\partial\Psi[\tilde{\mathbf{m}}]}{\partial r}
-l_{ex}^2\left[\Delta\tilde{m}^r+\left(\frac{1}{\sin^2\theta}-2\right)\frac{\tilde{m}^r}{r^2}
-\frac{2}{r^2}\left(\frac{\partial\tilde{m}^{\theta}}{\partial\theta}+\cot\theta\,\tilde{m}^\theta\right)\right]=\tilde{h}^{r}\,,\nonumber\\
&&\frac{1}{\omega_s}\frac{\partial\tilde{m}^{r}}{\partial t}+\frac{\alpha}{\omega_s}\frac{\partial\tilde{m}^{\theta}}{\partial t}
+K\tilde{m}^{\theta}+\frac{1}{r}\frac{\partial\Psi[\tilde{\mathbf{m}}]}{\partial\theta}
-l_{ex}^2\left[\Delta\tilde{m}^{\theta}+\frac{2}{r^2}\frac{\partial\tilde{m}^{r}}{\partial\theta}\right]=\tilde{h}^{\theta}\,,\label{LLlin}
\end{eqnarray}
\end{widetext}
where $\Delta$ is the Laplace operator, $\omega_s=\gamma M_s$, and
\begin{equation}\label{PsiSphericalTilde}
\Psi[\tilde{\mathbf{m}}]=\!\!\int\limits_{R_1}^{R_2}\!\!r'^2dr'\!\!\!\int\!\!d\Omega'
\left[\tilde{m}^{r}(\mathbf{r}')\frac{\partial}{\partial r'}+
\frac{\tilde{m}^{\theta}(\mathbf{r}')}{r'}\frac{\partial}{\partial\theta'}\right]\!\frac{1}{|\mathbf{r}-\mathbf{r}'|}\,.
\end{equation}
For the type-I magnetic anisotropy, we have to replace in Eq.~\eqref{LLlin}
\begin{eqnarray}
K\tilde{m}^r&\to&-K\left[\cos^2\!\theta\,\tilde{m}^r-\sin\theta\cos\theta\tilde{m}^{\theta}\right]\,,\nonumber\\
K\tilde{m}^{\theta}&\to&-K\left[\sin^2\!\theta\,\tilde{m}^{\theta}-\sin\theta\cos\theta\tilde{m}^{r}\right]\,.\label{LLlinSubst}
\end{eqnarray}
Note that for the considered equilibrium magnetic configuration~\eqref{M0}, the $\varphi$ component of magnetization dropped out from the equations~\eqref{LLlin}, we always have $\tilde{m}^{\varphi}=0$. The boundary conditions to this system of equations are obtained from Eq.~\eqref{LL0bound} by substitution $M^a=M_0^a+\tilde{m}^a/\varkappa^{a}$ and linearization by $\tilde{m}^a$. Since for the equilibrium configuration under consideration $M_0^{a;r}=0$, the boundary conditions are reduced to (c.f. with Refs.~\onlinecite{Brown1959,Aharoni1991})
\begin{equation}\label{LLbound}
\left.\frac{\partial\tilde{m}^{r}}{\partial r}\right|_{r=R_{1,2}}\!\!\!\!\!\!\!\!=0\,,\;\;
\left.\frac{\partial\tilde{m}^{\theta}}{\partial r}\right|_{r=R_{1,2}}\!\!\!\!\!\!\!\!=0\,.
\end{equation}
Thus, we derived the system of two linear integro-differential equations for the components $\tilde{m}^{r}$ and $\tilde{m}^{\theta}$. We will solve this system by expanding the magnetization components into a series of orthogonal functions
\begin{equation}\label{dirtrans}
\tilde{m}^{a}(\mathbf{r})=\sum_{k=0}^{\infty}\sum_{l=0}^{\infty}\sum_{m=-l}^{l}
\tilde{m}^{aklm}f_k(r)Y_l^{m}(\theta,\varphi)\,,
\end{equation}
where $Y_l^{m}(\theta,\varphi)$ are spherical harmonics. As for the radial functions, they should be orthogonal with weight $r^2$, i.e.,
\begin{equation}
\frac{1}{V}\int\limits_{R_1}^{R_2}\!r^2drf_k(r)f_{k'}(r)=\delta_{kk'}\,.
\end{equation}
They also should satisfy the boundary conditions
\begin{equation}\label{LLboundF}
\left.\frac{\partial f_k(r)}{\partial r}\right|_{r=R_{1,2}}\!\!\!\!\!\!\!\!=0\,,
\end{equation}
following from the conditions~\eqref{LLbound}. The algorithm of the construction of such functions is described in the Appendix. The transformation, inverse to Eq.~\eqref{dirtrans}, is
\begin{equation}\label{invtrans}
\tilde{m}^{aklm}=\frac{1}{V}\int\limits_{V}\!d^3\mathbf{r}\,\tilde{m}^{a}(\mathbf{r})f_k(r)Y_l^{m}(\theta,\varphi)^{*}\,.
\end{equation}
Substituting Eq.~\eqref{dirtrans} into Eq.~\eqref{LLlin}, multiplying both parts by $f_k(r)Y_l^{m}(\theta,\varphi){*}$ and integrating over the volume of the particle, after simple but cumbersome calculations, we obtain for the type-II magnetic anisotropy
\begin{widetext}
\begin{eqnarray}
&&-\frac{1}{\omega_s}\frac{\partial\tilde{m}^{\theta klm}}{\partial t}+\frac{\alpha}{\omega_s}\frac{\partial\tilde{m}^{rklm}}{\partial t}
+K\tilde{m}^{rklm}+
\sum_{k'=0}^{\infty}\left[\frac{4\pi F^{rr}_{kk'l}}{2l+1}+a_{ex}^2\left(G^{r}_{kk'}+(l^2+l+2)G^{0}_{kk'}\right)\right]\tilde{m}^{rk'lm}\nonumber\\
&&-(a^{(m)}_{ex})^2\sum_{k'=0}^{\infty}\sum_{l'=|m|}^{\infty}L_{ll'}^{(m)}(a_{ex})G^{0}_{kk'}\tilde{m}^{rk'l'm}
+\sum_{k'=0}^{\infty}\sum_{l'=|m|}^{\infty}\left[\frac{4\pi F^{r\theta}_{kk'l}}{2l+1}-
a^2_{ex}G^{0}_{kk'}\right]D^{(m)}_{ll'}\tilde{m}^{\theta k'l'm}=\tilde{h}^{rklm},\nonumber\\
&&\frac{1}{\omega_s}\frac{\partial\tilde{m}^{rklm}}{\partial t}+\frac{\alpha}{\omega_s}\frac{\partial\tilde{m}^{\theta klm}}{\partial t}
+K\tilde{m}^{\theta klm}+
\sum_{k'=0}^{\infty}\sum_{l'=|m|}^{\infty}\Bigg[\sum_{n=|m|}^{\infty}\frac{4\pi F^{\theta\theta}_{kk'n}}{2n+1}D^{(m)*}_{nl}D^{(m)}_{nl'}+
\nonumber\\
&&+a_{ex}^2\delta_{ll'}\left(G^{r}_{kk'}+l(l+1)G^{0}_{kk'}\right)\Bigg]\tilde{m}^{\theta k'l'm}
+\sum_{k'=0}^{\infty}\sum_{l'=|m|}^{\infty}\left[\frac{4\pi F^{\theta r}_{kk'l}}{2l'+1}-
a^2_{ex}G^{0}_{k'k}\right]D^{(m)*}_{l'l}\tilde{m}^{rk'l'm}=\tilde{h}^{\theta klm},\label{LLeqKLM}
\end{eqnarray}
\end{widetext}
where $a_{ex}=l_{ex}/R_0$, $[a^{(0)}_{ex}]^2=a^2_{ex}\ln(1/a_{ex})$, $a^{(m)}_{ex}=a_{ex}$ if $m\neq0$, $\tilde{h}^{aklm}$ are the components of the magnetic field decomposition defined similarly to Eq.~\eqref{invtrans}, and the remaining parameters not defined above are given below. Namely, we introduce the following notations:
\begin{equation}\label{Dmll}
D^{(m)}_{ll'}=\int d\Omega\,\frac{\partial Y_l^{m*}(\theta,\varphi)}{\partial\theta}Y_{l'}^{m}(\theta,\varphi)\,,
\end{equation}
\begin{equation}\label{Lll0}
L_{ll'}^{(0)}(a_{ex})=\sqrt{2l+1}\sqrt{2l'+1}\left|P_l^0(a_{ex})P_{l'}^0(a_{ex})\right|\,,
\end{equation}
\begin{eqnarray}
L_{ll'}^{(m)}(a_{ex})&=&\sqrt{\frac{(2l+1)(l-m)!}{(l+m)!}}\sqrt{\frac{(2l'+1)(l'-m)!}{(l'+m)!}}\times\nonumber\\
&&\int\limits_0^{1-a_{ex}}\!\!\!\!dx\,\frac{P_l^m(x)P_{l'}^m(x)}{1-x^2}\,,\;\;m\neq0\,,\label{Lllm}
\end{eqnarray}
where $P_l^m(x)$ are the associated Legendre polynomials,
\begin{equation}\label{G0}
G^0_{kk'}=\frac{R_0^2}{V}\int\limits_{R_1}^{R_2}dr\,f_k(r)f_{k'}(r),
\end{equation}
\begin{eqnarray}
G^r_{kk'}&=&\frac{R_0^2}{V}\Bigg\{\,\int\limits_{R_1}^{R_2}dr\,
\frac{\partial}{\partial r}\left[rf_k(r)\right]\frac{\partial}{\partial r}\left[rf_{k'}(r)\right]-\label{G0Gr}\\
&-&\left[R_2f_k(R_2)f_{k'}(R_2)-R_1f_k(R_1)f_{k'}(R_1)\right]\Bigg\},\nonumber
\end{eqnarray}
\begin{eqnarray}
F^{rr}_{kk'l}&=&\frac{1}{V}\,\int\limits_{R_1}^{R_2}r^2dr\,f_{k}(r)\frac{\partial}{\partial r}
\int\limits_{R_1}^{R_2}r'^2dr'\,\tau'_{l}(r,r')f_{k'}(r')\,,\nonumber\\
F^{\theta r}_{kk'l}&=&\frac{1}{V}\,\int\limits_{R_1}^{R_2}rdr\,f_{k}(r)
\int\limits_{R_1}^{R_2}r'^2dr'\,\tau'_{l}(r,r')f_{k'}(r')=F^{r\theta}_{k'kl}\,,\nonumber\\
F^{\theta\theta}_{kk'l}&=&\frac{1}{V}\,\int\limits_{R_1}^{R_2}rdr\,f_{k}(r)
\int\limits_{R_1}^{R_2}r'dr'\,\tau_{l}(r,r')f_{k'}(r')\,,\label{Fkkl}
\end{eqnarray}
where
\begin{equation}\label{tau}
\tau_{l}(r,r')=
\left\{\begin{array}{l}
\displaystyle\frac{r^l}{r'^{l+1}}\,,\;\;r<r'\\
\displaystyle\frac{r'^l}{r^{l+1}}\,,\;\;r>r'
\end{array}\right.\,,\;\;\tau'_{l}(r,r')=\frac{\partial\tau_{l}(r,r')}{\partial r'}\,.
\end{equation}
Deriving Eq.~\eqref{LLeqKLM} (in the part concerning the magnitostatic energy) we used the following relation
\begin{equation}\label{exp}
\frac{1}{|\mathbf{r}-\mathbf{r}'|}=4\pi\sum_{l=0}^{\infty}\frac{\tau_{l}(r,r')}{2l+1}
\sum_{m=-l}^{l}Y_l^{m}(\theta,\varphi)Y_l^{m*}(\theta',\varphi')\,.
\end{equation}
Note that the presence of the factor $\ln(1/a_{ex})$ in the definition of $[a^{(0)}_{ex}]^2$, as well as the fact that we restrict integration over $x$ in the formula~\eqref{Lllm} (despite the fact that the integral is convergent) is due to the presence of the vortex cores located at the poles of the particle. We neglect the effect of these cores. In Eq.~\eqref{LLeqKLM}, the terms proportional to $F^{ab}_{kk'l}$ come from the magnetostatic energy of the particle. They play the role of demagnetizing factors. Terms proportional to $a_{ex}^2$ or $(a^{(m)}_{ex})^2$ arise due to the exchange energy. For relatively small $k$ and $l$, these terms are small compared to the ``demagnetizing factors'', since in the case under consideration $a_{ex}\sim10^{-2}$. This limit is different from that considered, e.g., in Refs.~\onlinecite{Aharoni1991,Aharoni1997,viau1997size}. At the same time, for large $k$ or $l$, the contributions from the exchange energy become large (these are proportional to $k^2$ and $l^2$, respectively). This ensures that Eq.~\eqref{LLeqKLM} can be solved by restriction of the summation over $k$ and $l$ to some finite values, if we are interested in not too high frequencies.

Formula~\eqref{LLeqKLM} is derived for the type-II magnetic anisotropy. For the type-I magnetic anisotropy we have to replace in Eq.~\eqref{LLeqKLM}
\begin{eqnarray}
K\tilde{m}^{rlkm}&\to&-\frac{K}{2}\left[\tilde{m}^{rlkm}+\sum_{l'=|m|}^{\infty}C_{ll'}^{(m)}\tilde{m}^{rlkm}\right.\nonumber\\
&&\left.-\sum_{l'=|m|}^{\infty}S_{ll'}^{(m)}\tilde{m}^{\theta lkm}\right],\nonumber\\
K\tilde{m}^{\theta lkm}&\to&-\frac{K}{2}\left[\tilde{m}^{\theta lkm}-\sum_{l'=|m|}^{\infty}C_{ll'}^{(m)}\tilde{m}^{\theta lkm}\right.\nonumber\\
&&\left.-\sum_{l'=|m|}^{\infty}S_{ll'}^{(m)}\tilde{m}^{rlkm}\right],\label{LLeqKLMSubst}
\end{eqnarray}
where
\begin{eqnarray}
C_{ll'}^{(m)}&=&\int\!\!d\Omega\,Y_l^{m*}(\theta,\varphi)\cos(2\theta)Y_{l'}^{m}(\theta,\varphi)\,,\nonumber\\
S_{ll'}^{(m)}&=&\int\!\!d\Omega\,Y_l^{m*}(\theta,\varphi)\sin(2\theta)Y_{l'}^{m}(\theta,\varphi)\,.\label{CS}
\end{eqnarray}

Thus, we obtained a system of linear equations for the components of magnetization $\tilde{m}^{aklm}$, connecting them with the components of the external alternating magnetic field $\tilde{h}^{aklm}$. An important property of these equations is that they are diagonal in the azimuth index $m$. This is directly related to the geometry of the problem under consideration. In arbitrarily directed, homogeneous external field~\cite{Note2}, only the components $\tilde{h}^{aklm}$ with $m=0$ and $m=\pm1$ are nonzero. Accordingly, only the magnetization components $\tilde{m}^{aklm}$ with the same indices $m$ will be excited. We solve the system~\eqref{LLeqKLM} numerically by truncating the summation over $k$ and $l$ to some values of $K_{\text{max}}-1$ and $L_{\text{max}}-1$, respectively. Let us introduce the vectors $\tilde{m}_{m}$ and $\tilde{h}_{m}$, having dimension $2K_{\text{max}}L_{\text{max}}$, in the form
\begin{eqnarray}
\tilde{m}_{m}&=&\left(\begin{array}{c}
\tilde{m}_{m}^{r}\\ \tilde{m}_{m}^{\theta}
\end{array}\right)=
\left(\begin{array}{c}
\{\tilde{m}^{rklm}\}\\ \{\tilde{m}^{\theta klm}\}
\end{array}\right),\\
\tilde{h}_{m}&=&\left(\begin{array}{c}
\tilde{h}_{m}^{r}\\ \tilde{h}_{m}^{\theta}
\end{array}\right)=
\left(\begin{array}{c}
\{\tilde{h}^{rklm}\}\\ \{\tilde{h}^{\theta klm}\}
\end{array}\right).
\end{eqnarray}
Then, in the frequency representation, Eq.~\eqref{LLeqKLM} can be rewritten in the matrix form as
\begin{equation}\label{LLeqMatrix}
\left[-\frac{i\omega}{\omega_s}\left(\begin{array}{cc}
\alpha&-1\\
1&\alpha
\end{array}\right)+
\left(\!\begin{array}{cc}
\hat{\Lambda}_m^{rr}&(\hat{\Lambda}_m^{\theta r})^T\\
\hat{\Lambda}_m^{\theta r}&\hat{\Lambda}_m^{\theta\theta}
\end{array}\!\right)\right]\!\!\left(\begin{array}{c}
\tilde{m}_{m}^{r}\\ \tilde{m}_{m}^{\theta}
\end{array}\right)\!=\!
\left(\begin{array}{c}
\tilde{h}_{m}^{r}\\ \tilde{h}_{m}^{\theta}
\end{array}\right)\!,
\end{equation}
where elements of the matrices $\hat{\Lambda}_m^{ab}$ are obviously determined from Eq.~\eqref{LLeqKLM}. It is not difficult to verify that the matrices $\hat{\Lambda}_m^{ab}$ are real, and the matrices $\hat{\Lambda}_m^{aa}$ are symmetric. Let us denote the matrix in square brackets in Eq.~\eqref{LLeqMatrix} as $\hat{\tilde{\chi}}_{m}^{-1}(\omega)$. The inverse matrices with $m=0,\,\pm1$, the elements of which we denote as $\tilde{\chi}^{aklm}_{bk'm}(\omega)$, determine the susceptibility of the particle in the chosen representation. In the next Section, we will derive the corresponding expression for the susceptibility of a particle in the Cartesian coordinate system, as well as an expression for the magnetic permeability of the composite as a function of $\tilde{\chi}^{aklm}_{bk''m}(\omega)$.

\section{Permeability of the composite}\label{Permeability}

\begin{figure}[t]
\centering
\includegraphics[width=0.99\columnwidth]{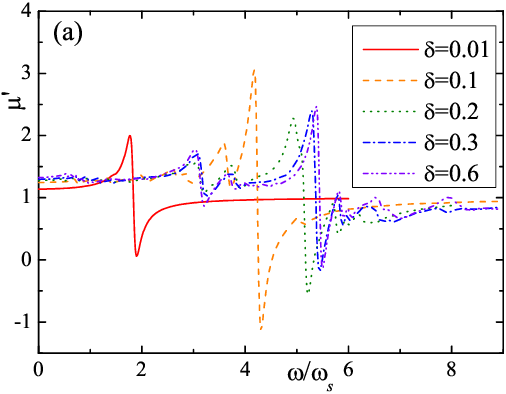}\\
\includegraphics[width=0.99\columnwidth]{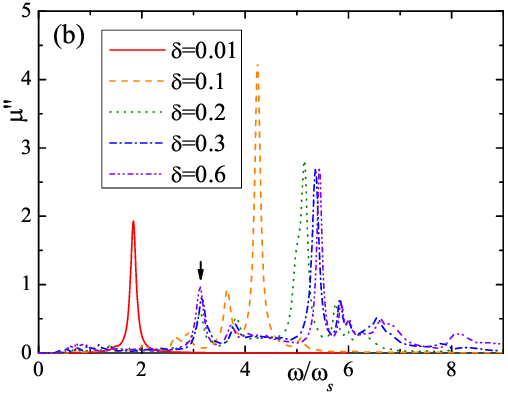}
\caption{\label{FigMuvsDelta} Frequency dependencies of real (a) and imaginary (b) parts of the permeability of the composite, calculated for five values (see the legend) of the shell thickness $d$. Model parameters: $K=0.1$ (type-II), $a_{ex}=0.01$, $p_c=0.15$, $\alpha=0.01$.}
\end{figure}

In Cartesian coordinates and frequency representation, the relationship between the magnetization of a particle and an external field is defined as
\begin{equation}\label{LLchirr}
m^{\alpha}(\mathbf{r};\omega)=\frac{1}{V}\!\!\int\limits_{V}\!\!d^3\mathbf{r}'\,\chi^{\alpha}_{\beta}(\mathbf{r},\mathbf{r}';\omega)h^{\beta}(\mathbf{r}';\omega)\,,
\end{equation}
where $\chi^{\alpha}_{\beta}(\mathbf{r},\mathbf{r}';\omega)$ is the particle susceptibility tensor in Cartesian coordinate system. It is related to that in spherical coordinate system by the equality
\begin{equation}\label{chitochi}
\chi^{\alpha}_{\beta}(\mathbf{r},\mathbf{r}';\omega)=\frac{\partial x^{\alpha}}{\partial x^{a}}
\chi^{a}_{b}(\mathbf{r},\mathbf{r}';\omega)\frac{\partial x'^{b}}{\partial x'^{\beta}}\,.
\end{equation}
In a homogeneous field, the averaged magnetization of a particle is
\begin{equation}\label{LLchirrav}
m^{\alpha}(\omega)=\frac{1}{V^2}\!\!\int\limits_{V}\!\!d^3\mathbf{r}\!\!\int\limits_{V}\!\!d^3\mathbf{r}'
\,\chi^{\alpha}_{\beta}(\mathbf{r},\mathbf{r}';\omega)\,h^{\beta}(\omega)\equiv\chi^{\alpha}_{\beta}(\omega)h^{\beta}(\omega)\,.
\end{equation}
We assume that the directions of the local $z$ axes of the particles are randomly distributed in the composite. Then, for the susceptibility of the composite, we obtain
\begin{equation}\label{chicomp}
\chi(\omega)=\frac{p_F}{3}\chi^{\alpha}_{\alpha}(\omega)=
\frac{p_F}{3V^2}\!\!\int\limits_{V}\!\!d^3\mathbf{r}\!\!\int\limits_{V}\!\!d^3\mathbf{r}'
\frac{\partial x^{\alpha}}{\partial x^{a}}\chi^{a}_{b}(\mathbf{r},\mathbf{r}';\omega)\frac{\partial x'^{b}}{\partial x'^{\alpha}}\,,
\end{equation}
where $p_F$ is the volume fraction of the ferromagnet. It is related to the volume fraction of particles in the composite, $p_c$, by the formula
\begin{equation}\label{pF}
p_F=p_c\left[1-\left(\frac{1-\delta}{1+\delta}\right)^3\right],\;\;\delta=\frac{R_2-R_1}{R_2+R_1}=\frac{2d}{R_0}.
\end{equation}
Note that the formula~\eqref{chicomp} was obtained by neglecting the interaction of particles with each other. As we have shown above, particles in the equilibrium configuration do not create a magnetic field (if vortex cores are ignored). Therefore, it can be expected that the approximation of non-interacting particles is adequate even for sufficiently large $p_c$. Particle susceptibility tensor in spherical coordinates, $\chi^{a}_{b}(\mathbf{r},\mathbf{r}';\omega)$, is related to the matrices we obtained above $\tilde{\chi}^{aklm}_{bk'l'm}(\omega)$ through the equality
\begin{eqnarray}
\chi^{a}_{b}(\mathbf{r},\mathbf{r}';\omega)&=&\frac{1}{\varkappa^{a}(r)}\Bigg[\sum_{m=-1}^{1}\sum_{kk'=0}^{\infty}\sum_{ll'=|m|}^{\infty}
f_k(r)Y_{l}^{m}(\theta,\varphi)\times\nonumber\\
&&\tilde{\chi}^{aklm}_{bk'l'm}(\omega)f_{k'}(r')Y_{l'}^{m}(\theta',\varphi')^{*}\Bigg]\varkappa^{b}(r')\,.\label{chipart}
\end{eqnarray}
Substituting this expression into Eq.~\eqref{chicomp} and integrating over the coordinates, we obtain for the magnetic permeability of the composite, $\mu(\omega)=1+4\pi\chi(\omega)$, the formula
\begin{widetext}
\begin{equation}\label{mucomp}
\mu(\omega)=1+\frac{4\pi p_F}{9}\!\!\!\!\sum_{m=-1}^{1}\!\!\left[
\tilde{\chi}^{r01m}_{r01m}(\omega)+\sum_{l=|m|}^{\infty}\left(\tilde{\chi}^{r01m}_{\theta0lm}(\omega)+
\tilde{\chi}^{\theta0lm}_{r01m}(\omega)\right)D^{(m)}_{1l}+
\sum_{l=|m|}^{\infty}\sum_{l'=|m|}^{\infty}\tilde{\chi}^{\theta0lm}_{\theta0l'm}(\omega)D^{(m)}_{1l}D^{(m)}_{1l'}\right]\!,
\end{equation}
\end{widetext}
where $D^{(m)}_{1l}$ is given by Eq.~\eqref{Dmll}. Deriving formula~\eqref{mucomp}, we took into account that for the selected set of functions $f_k(r)$ we have (see Appendix) $f_0(r)=\sqrt{4\pi}=\text{const}$, and
\begin{equation}\label{gk}
\frac{1}{V}\,\int\limits_{R_1}^{R_2}r^2dr\,f_{k}(r)=0\,,\;\;k>0\,,
\end{equation}
due to the orthogonality of the radial functions.

For a given frequency $\omega$ we calculate $\tilde{\chi}^{aklm}_{bk'm}(\omega)$ and $\mu(\omega)$ numerically. For all the results below, wherever it is not specified in the text, we choose $K_{\text{max}}=36$ and $L_{\text{max}}=151$. We first present the results for the type-II magnetic anisotropy. Figure~\ref{FigMuvsDelta} shows the frequency dependencies of the permeability of the composite, calculated for several values of the shell thickness $d$. The imaginary part of the permeability is characterized by one main peak and several secondary peaks, which are located at both lower and higher frequencies. When parameter $\delta=2d/R_0$ is small enough ($\delta\lesssim0.3$) the main peak shifts to the right as $\delta$ increases. This correlates with the results obtained in Ref.~\onlinecite{VortexConfiguration}. Such a behavior can be explained as following. For small $\delta$ one can consider only one, homogeneous, radial mode, taking $K_{\text{max}}=1$. The parameters $F^{ab}_{00l}$ can be calculated analytically. We will not present corresponding formulas here but will limit ourselves to the asymptotics at $\delta\to0$ and $l\to\infty$. We have
\begin{eqnarray}
F^{rr}_{00l}&\to&\left\{\begin{array}{l}
2l+1,\;\;\delta\to0\\
1/\delta,\;\;l\to\infty
\end{array}\right.,\;\;
F^{r\theta}_{00l}\to\left\{\begin{array}{l}
-\delta,\;\;\delta\to0\\
-1/l,\;\;l\to\infty
\end{array}\right.,\nonumber\\
F^{\theta\theta}_{00l}&\to&\left\{\begin{array}{l}
2\delta,\;\;\delta\to0\\
2/l,\;\;l\to\infty
\end{array}\right..\label{Fas}
\end{eqnarray}
Note that the asymptotics for $\delta\to0$ are valid only when $l\delta\ll1$. We see that the parameters $F^{r\theta}_{00l}$ and $F^{\theta\theta}_{00l}$ increase in magnitude when $\delta$ increases, which leads to a shift in the resonant frequency to the right. For larger $\delta(\gtrsim0.3)$, the position of the main resonance practically does not change. Indeed, the main resonance frequency cannot go to infinity when $\delta\to1$. When $\delta\lesssim0.1$, the amplitude of the main resonance peak increases with $\delta$. This is because the volume fraction of the ferromagnet, $p_F$, almost linearly increases with $\delta$ when $\delta$ is small. For larger $\delta$, the $p_F$ increases more slowly. Thus, the amplitude of the main resonance start to decrease when the position of the main peak shifts to higher frequencies. Let us estimate the characteristic values of the main resonance frequencies for the curves shown in Fig.~\ref{FigMuvsDelta}. Taking $M_s=1700$\,Gs (as for Fe), we will have $\omega_s/2\pi\cong4.75$\,GHz. Thus, the resonance frequencies lay approximately in the range $8$\,--\,$25$\,GHz. These resonance frequencies are bit higher than that observed experimentally in Refs.~\onlinecite{coatings10100995,s22083086,magnetism3020008}.

\begin{widetext}
\begin{figure*}[t]
\centering
\includegraphics[width=0.19\textwidth]{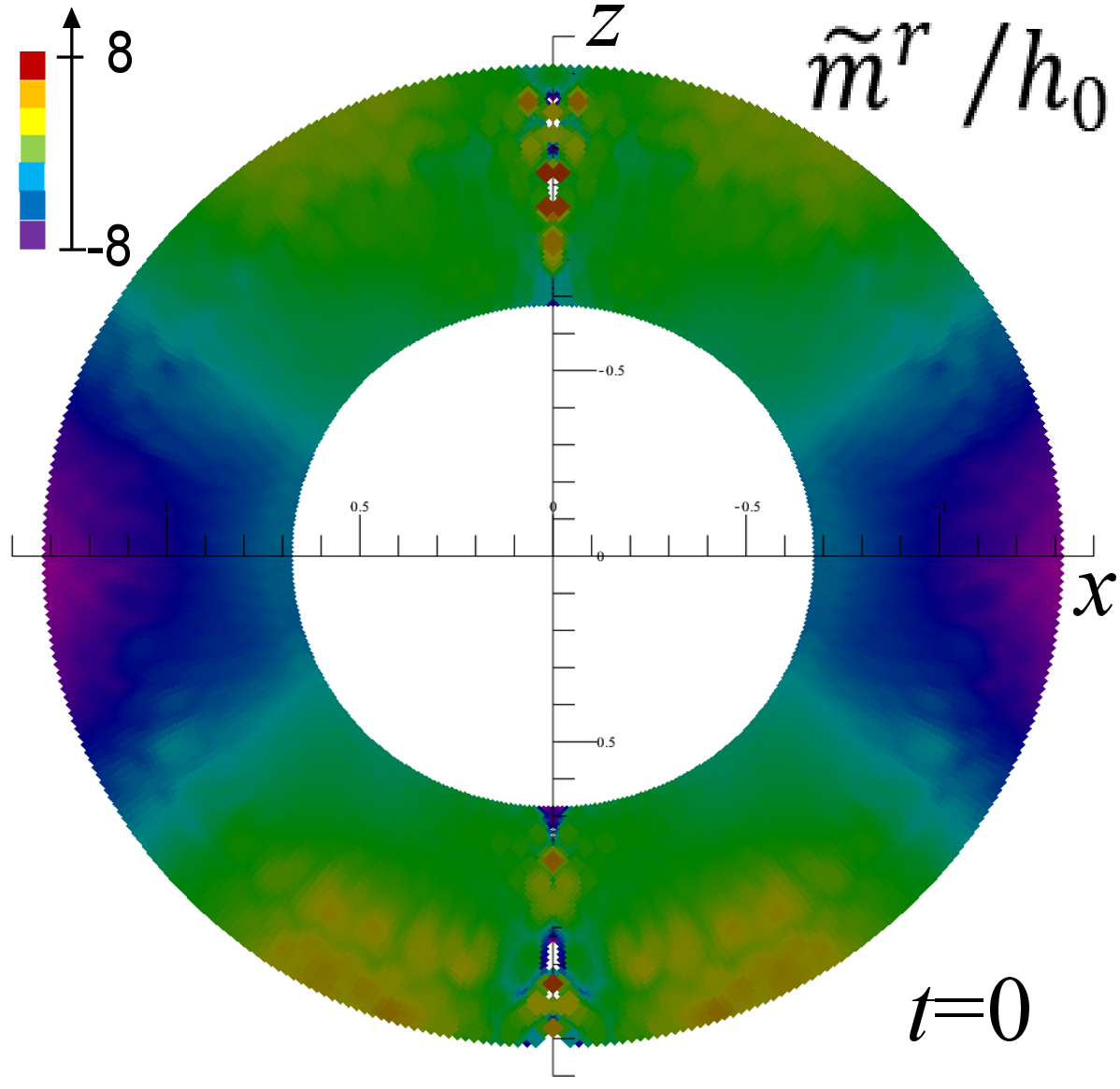}
\includegraphics[width=0.19\textwidth]{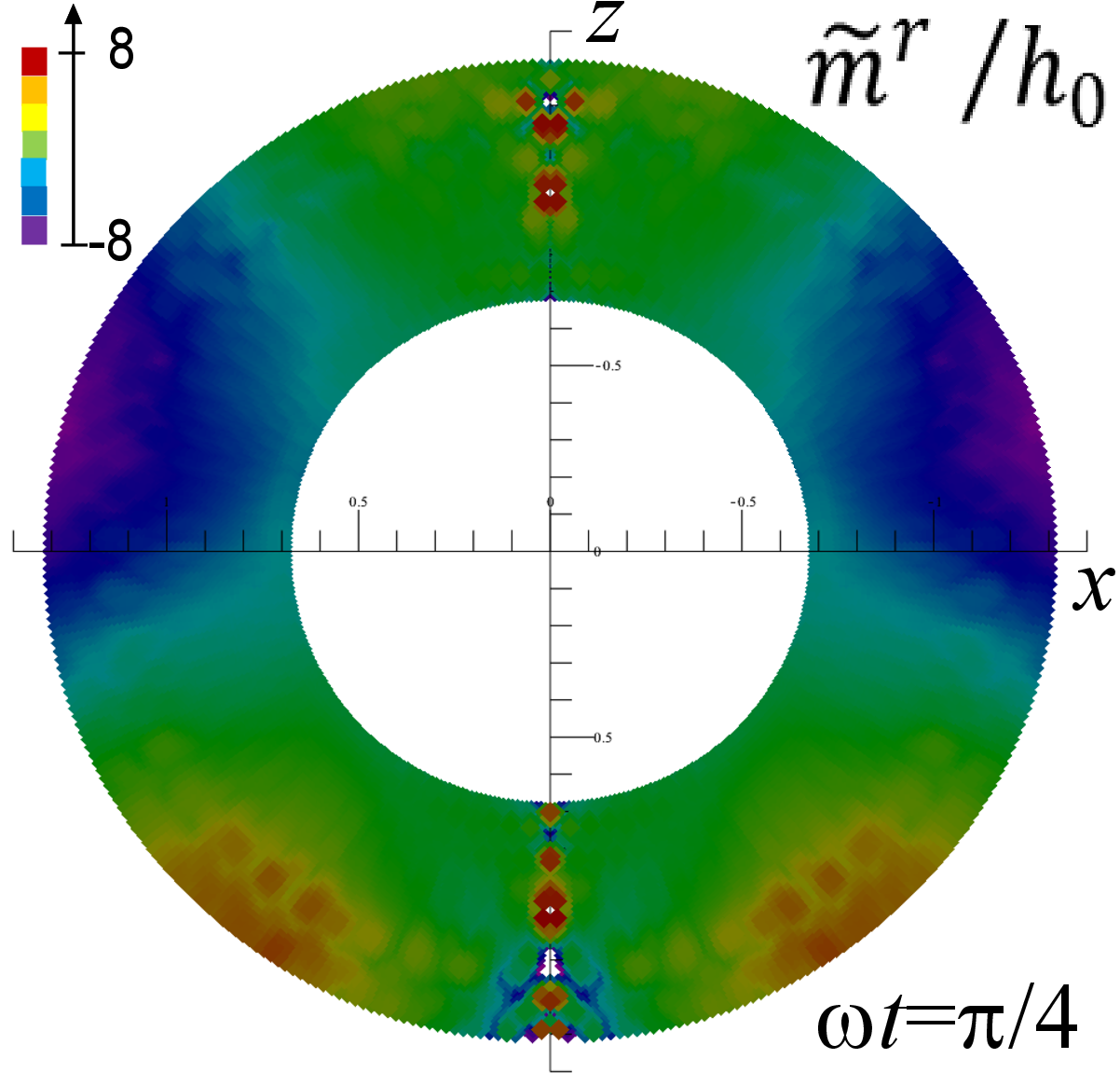}
\includegraphics[width=0.19\textwidth]{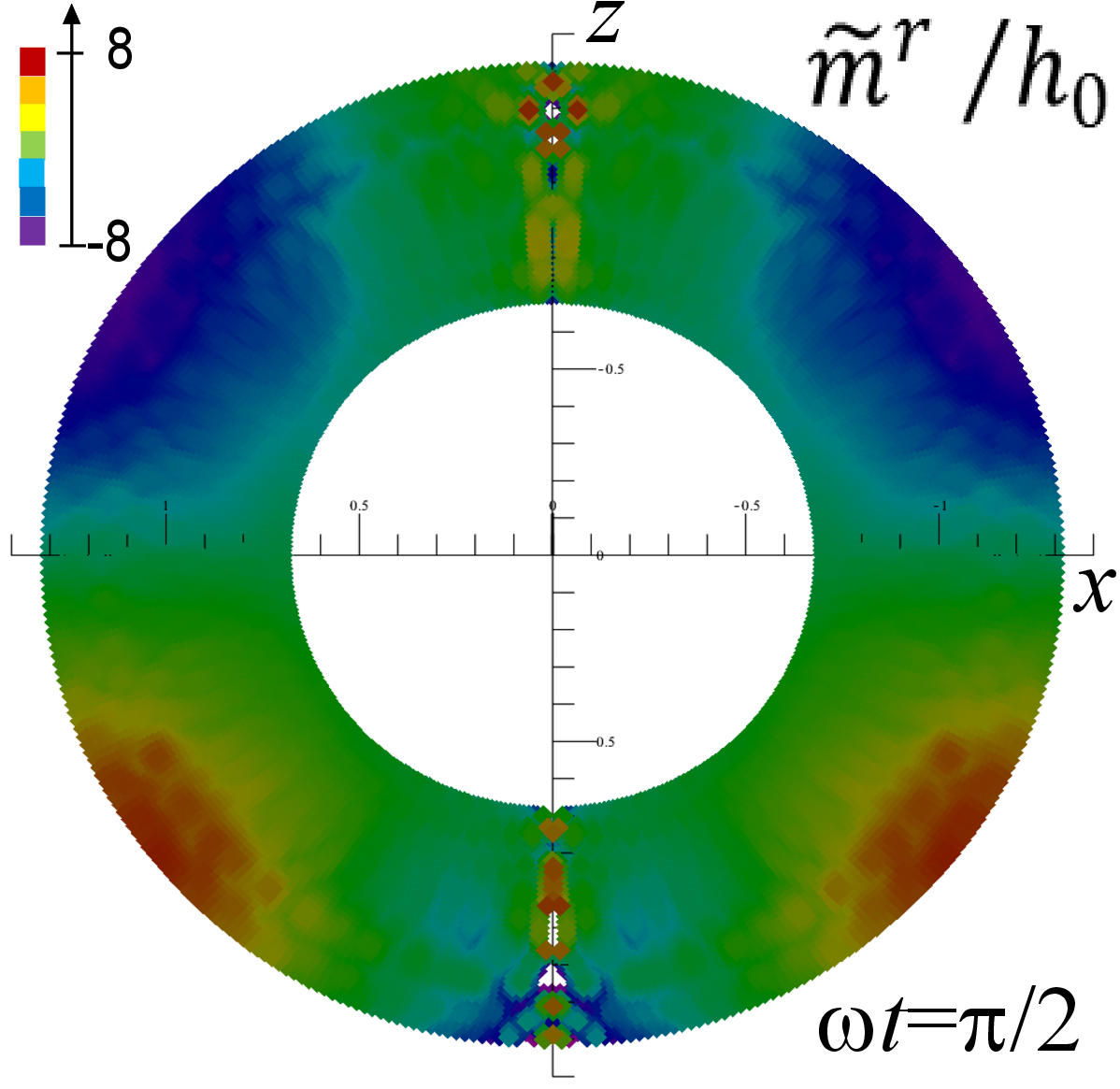}
\includegraphics[width=0.19\textwidth]{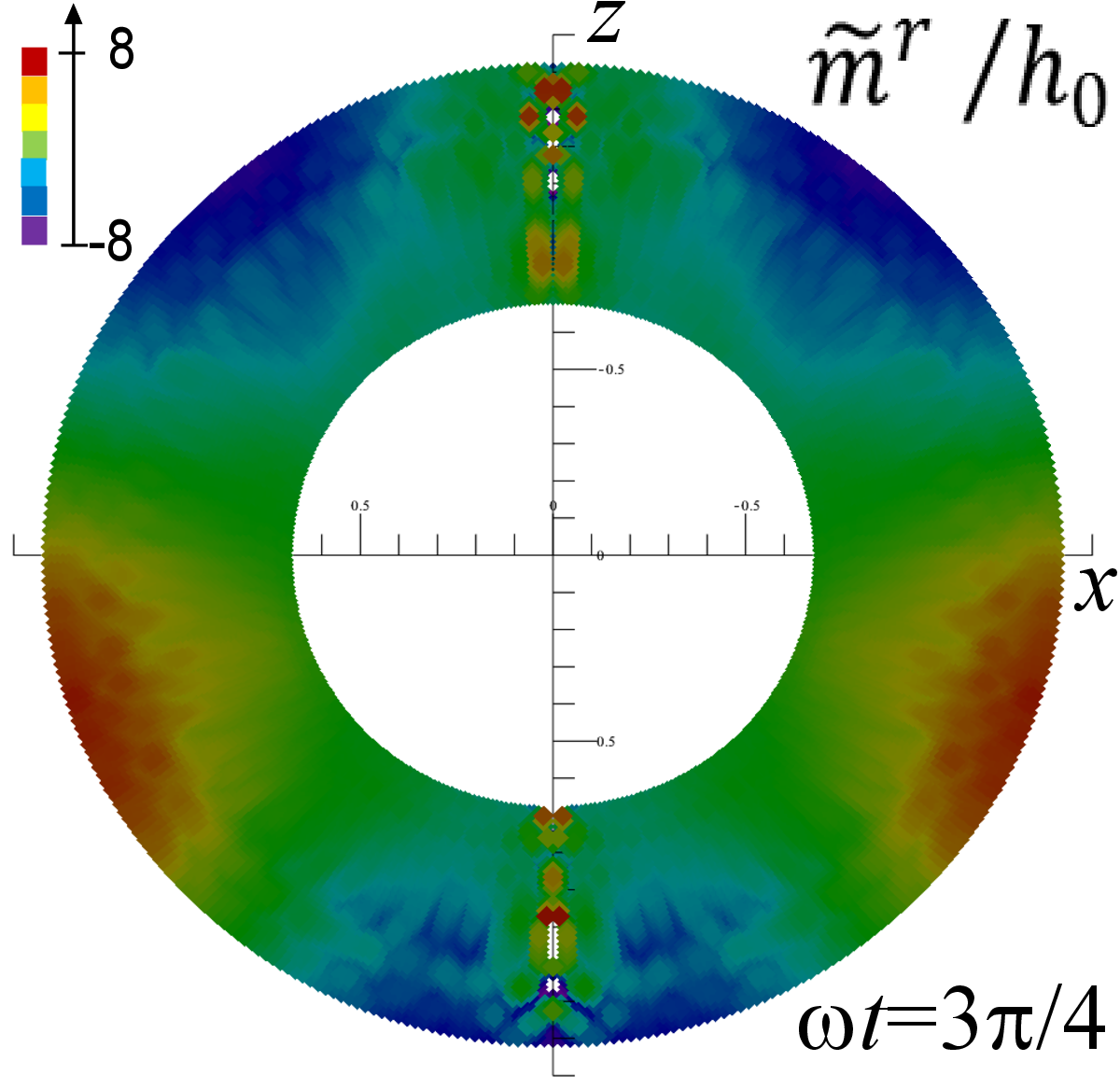}
\includegraphics[width=0.19\textwidth]{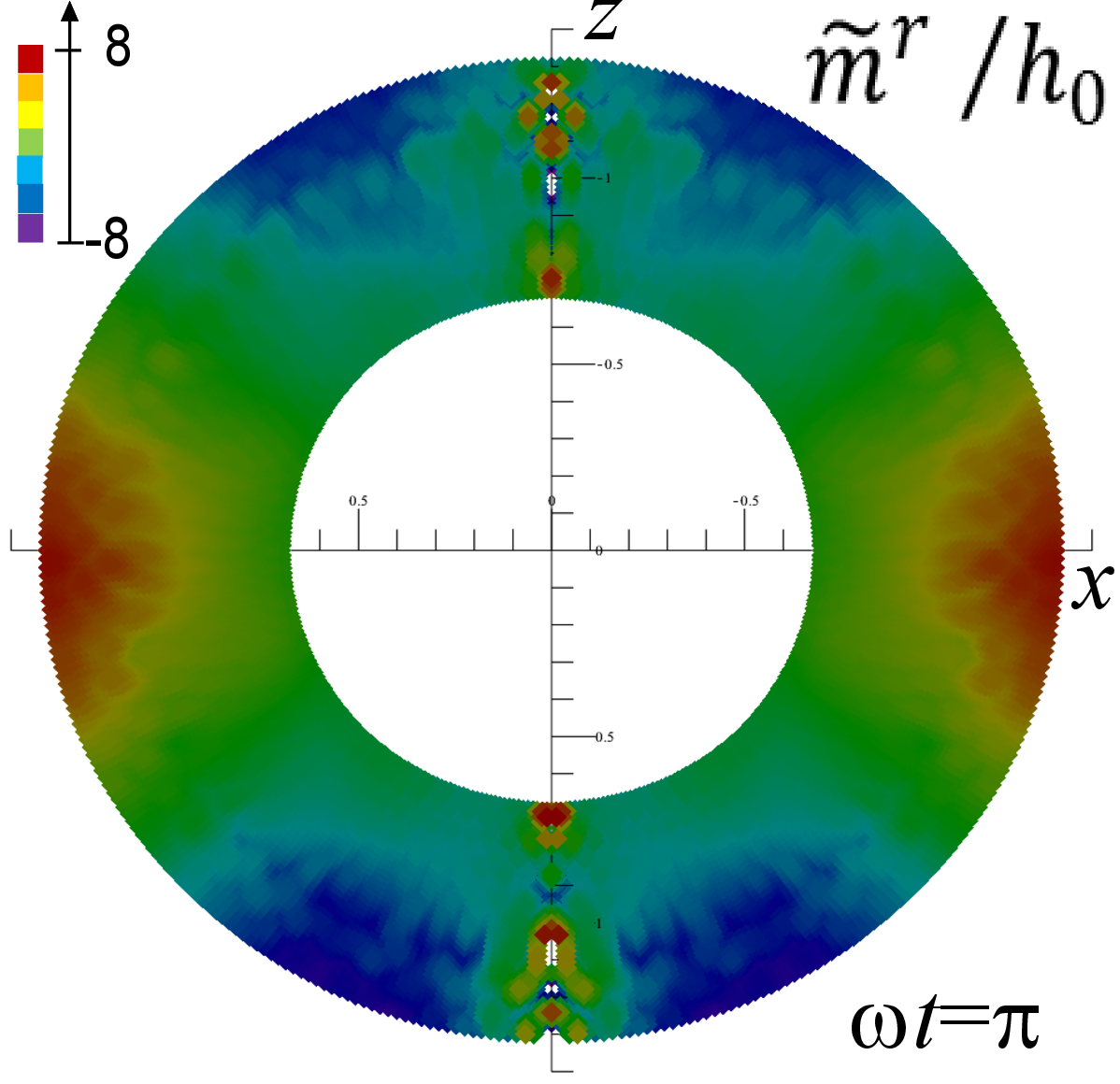}\vspace{5mm}\\
\includegraphics[width=0.19\textwidth]{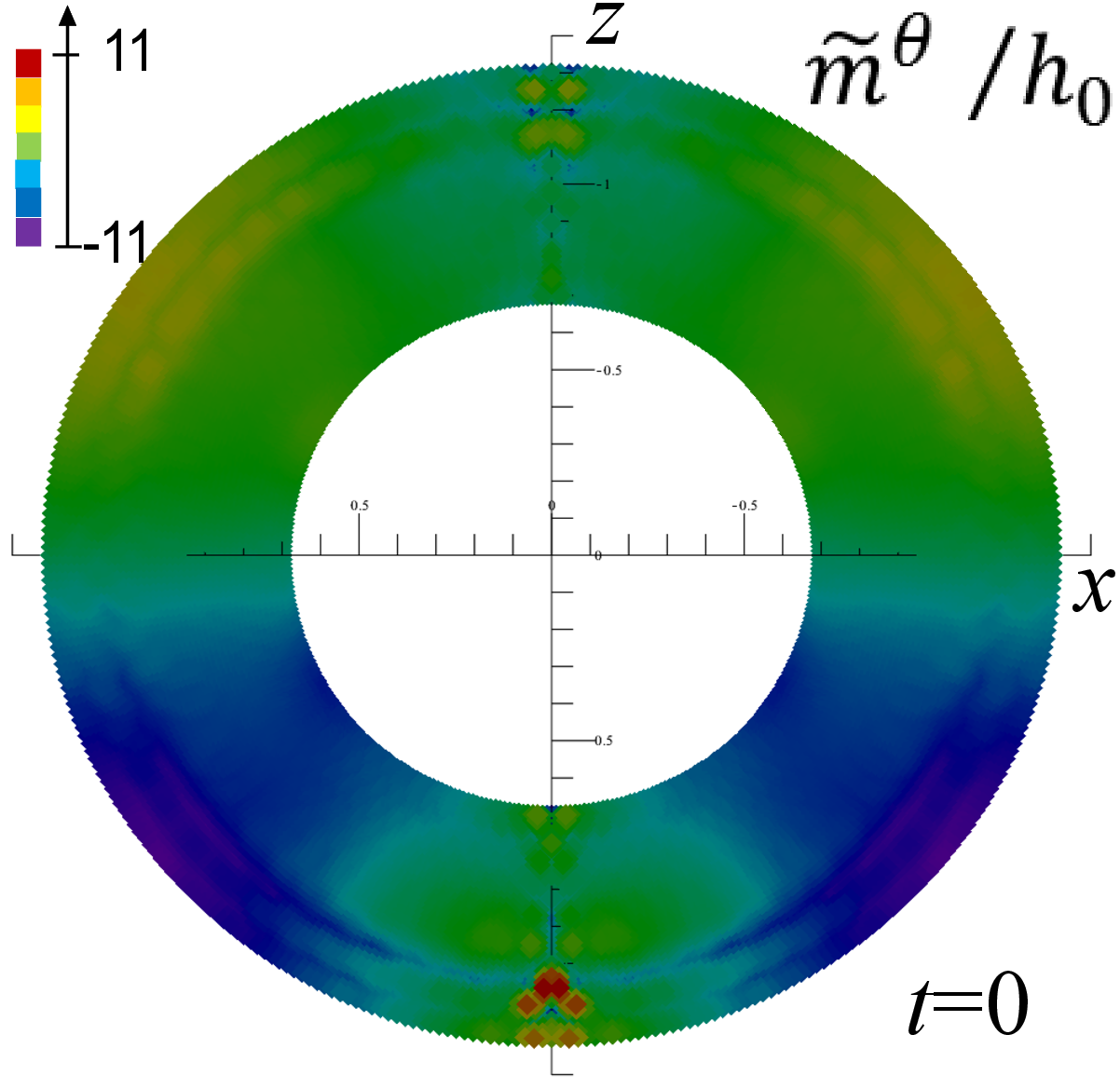}
\includegraphics[width=0.19\textwidth]{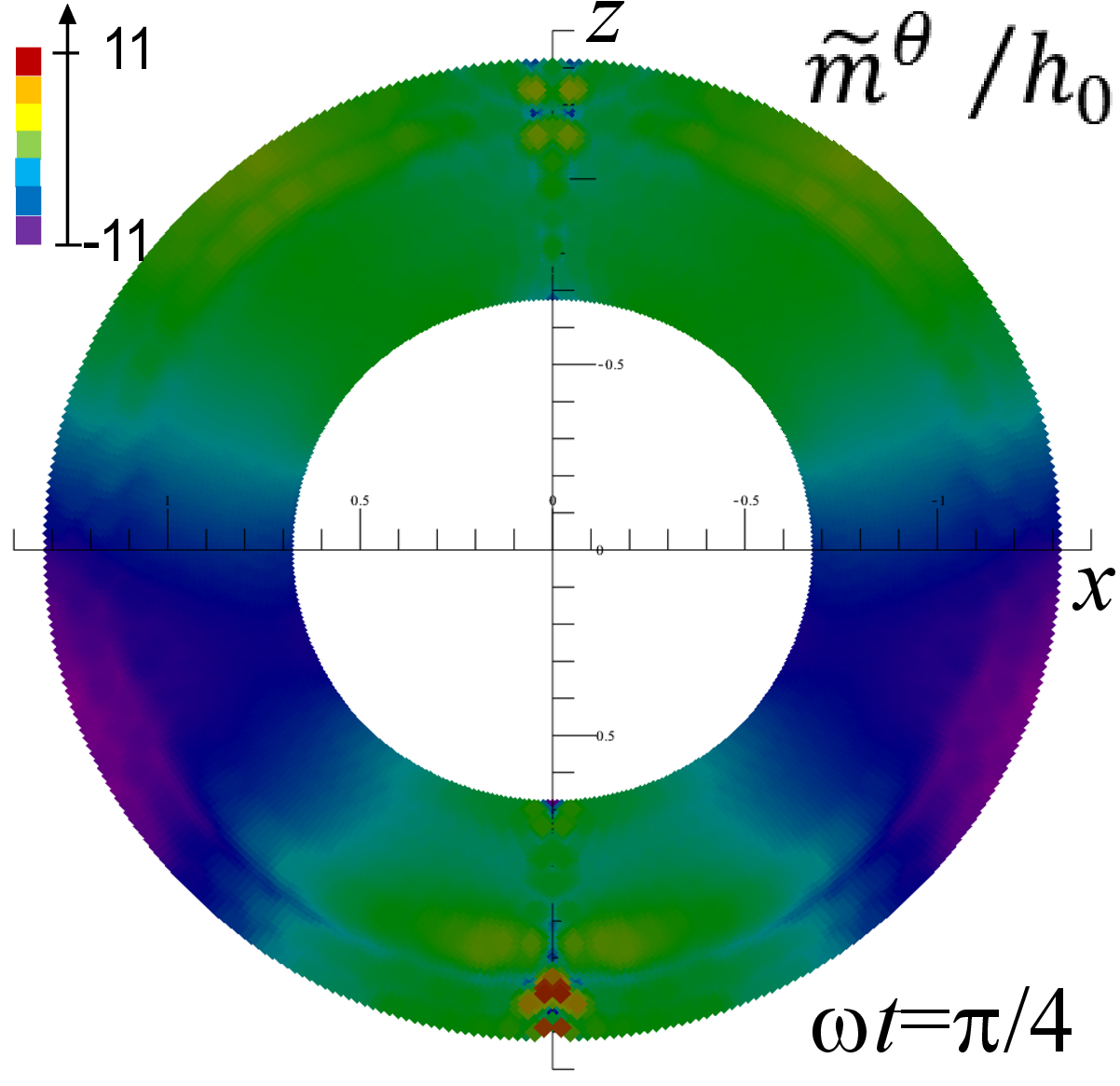}
\includegraphics[width=0.19\textwidth]{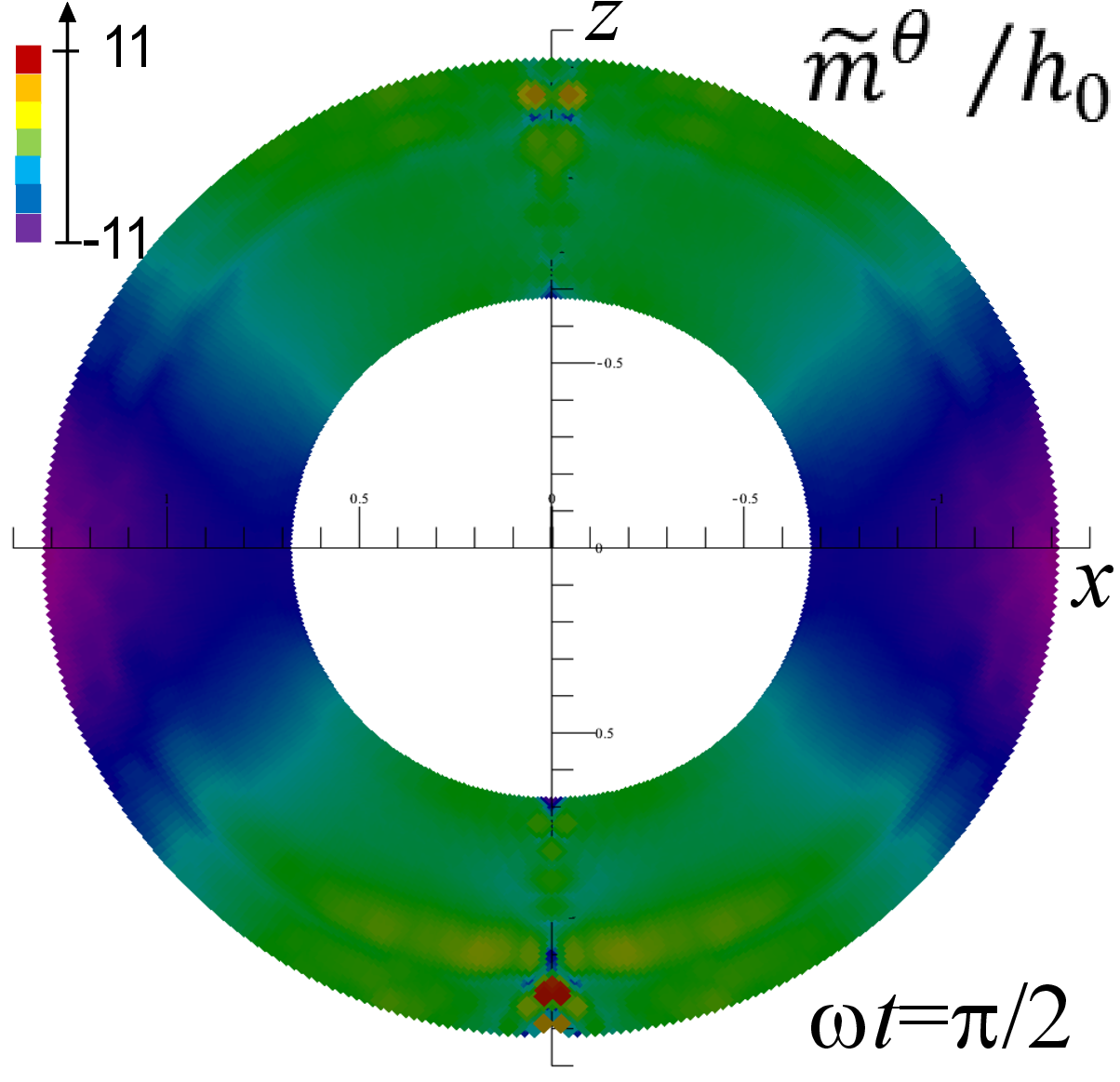}
\includegraphics[width=0.19\textwidth]{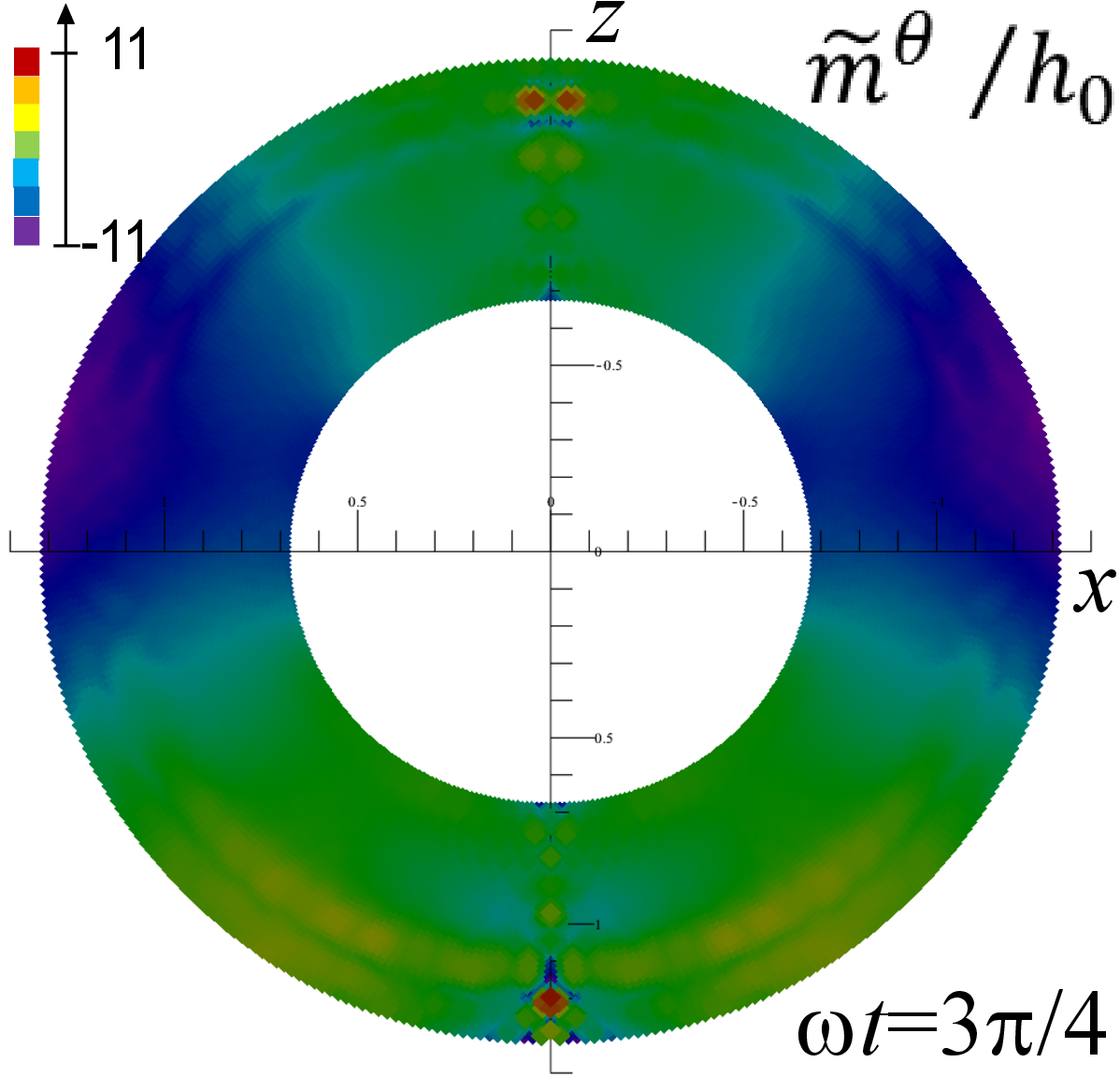}
\includegraphics[width=0.19\textwidth]{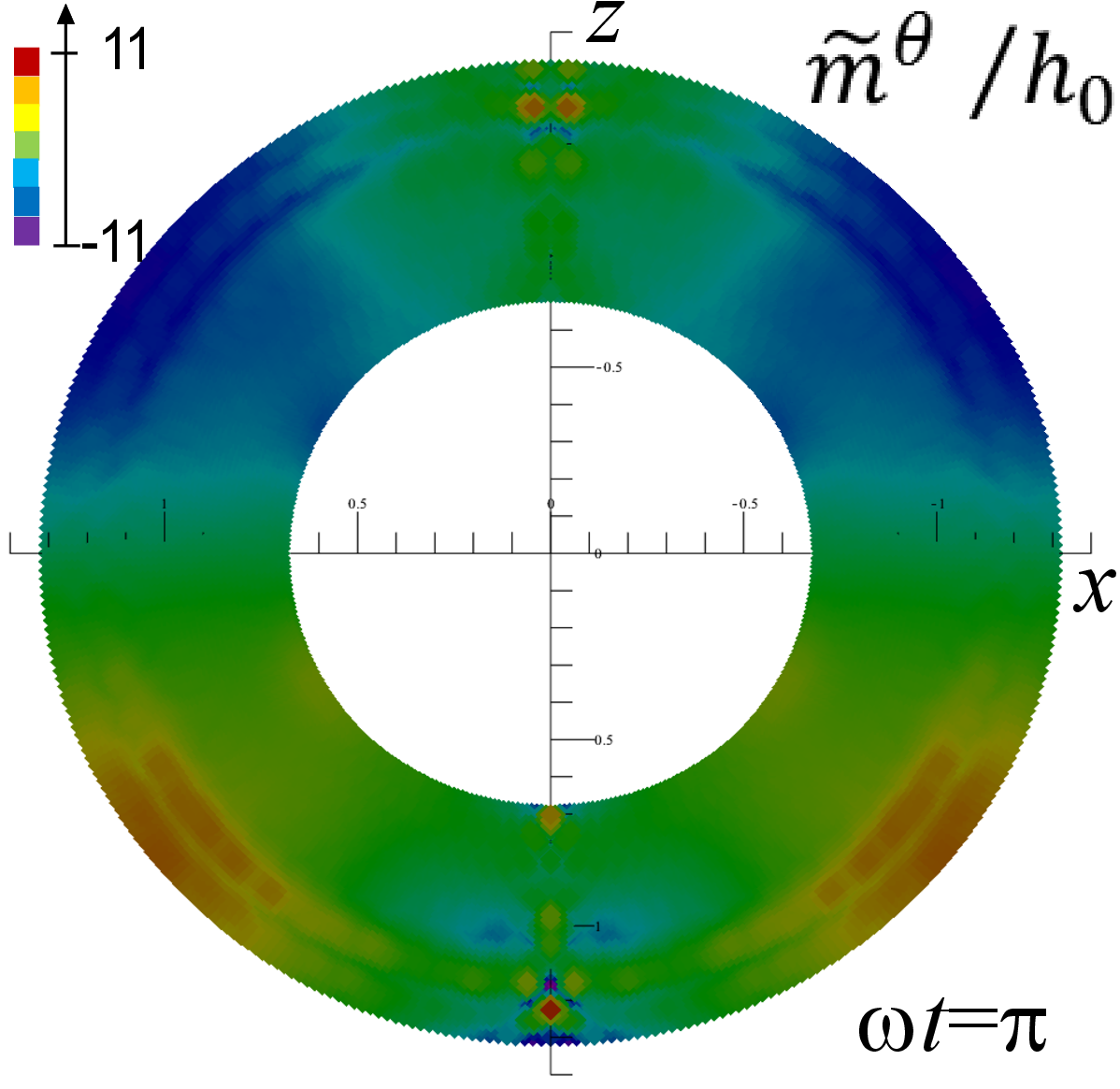}
\caption{\label{FigM1} The color plots of the radial (upper panels) and polar (lower panels) components of the particle's magnetization in the $xz$ plane, calculated at different $t$ in the range $0\leq\omega t\leq\pi$. The oscillation frequency corresponds to the main resonance. External ac magnetic field is parallel to the $z$ axis. Model parameters: $\delta=0.3$, $K=0.1$ (type-II), $a_{ex}=0.01$, $\alpha=0.01$.}
\end{figure*}
\end{widetext}

Let us analyze the character of the magnetization oscillations at different frequencies. We consider first the situation when the magnetic field is directed along the $z$ axis of a particle, i.e. perpendicular to $\mathbf{M}_0(\mathbf{r})$ (outside the vortex core). Using the expression~\eqref{LLchirr}, as well as the formula~\eqref{chipart} for the susceptibility of the particle, we obtain the formula for the particle's magnetization in the field $\mathbf{h}=\mathbf{e}_{z}h_0e^{-i\omega t}$ in the form
\begin{eqnarray}\label{mprec}
\tilde{m}^{a}(\mathbf{r},\,t)&=&\Real\left\{\sum_{k=0}^{\infty}\sum_{l=0}^{\infty}f_k(r)\sqrt{\frac{2l+1}{12\pi}}P_{l}(\cos\theta)\times\right.\\
&&\left.\left[\tilde{\chi}^{akl0}_{r010}(\omega)+\sum_{l'=0}^{\infty}\tilde{\chi}^{akl0}_{\theta0l'0}(\omega)D^{(0)}_{1l'}\right]e^{-i\omega t}\right\}h_0\,.\nonumber
\end{eqnarray}
We see, that the magnetization does not depend on the angle $\varphi$. For the chosen direction of the magnetic field the magnetization osculations are determined by the susceptibility matrix $\tilde{\chi}^{aklm}_{bk'l'm}(\omega)$ with $m=0$. Figure~\ref{FigM1} shows the snapshots of the $\tilde{m}^{a}(\mathbf{r},\,t)$ in the $xz$ plane taken at five different $t$ in the range $0\leq\omega t\leq\pi$. The frequency $\omega$ corresponds to the main resonance (the highest peak in Fig.~\ref{FigMuvsDelta}). The distribution of $\tilde{m}^{a}(\mathbf{r},\,t)$ is spatially no-uniform. Neglecting the sharp peaks at the poles, the largest magnitude of oscillations exists at the equator of the particle. The spatial profiles of the magnetization components $\tilde{m}^{r}(\theta,\,t)$ and $\tilde{m}^{\theta}(\theta,\,t)$ are phase shifted by $\pi/2$ relative to each other, which corresponds to the rotation of the local magnetization around an axis lying in the $xy$ plane and tangent to the surface of the particle.

\begin{figure}[t]
\centering
\includegraphics[width=0.49\columnwidth]{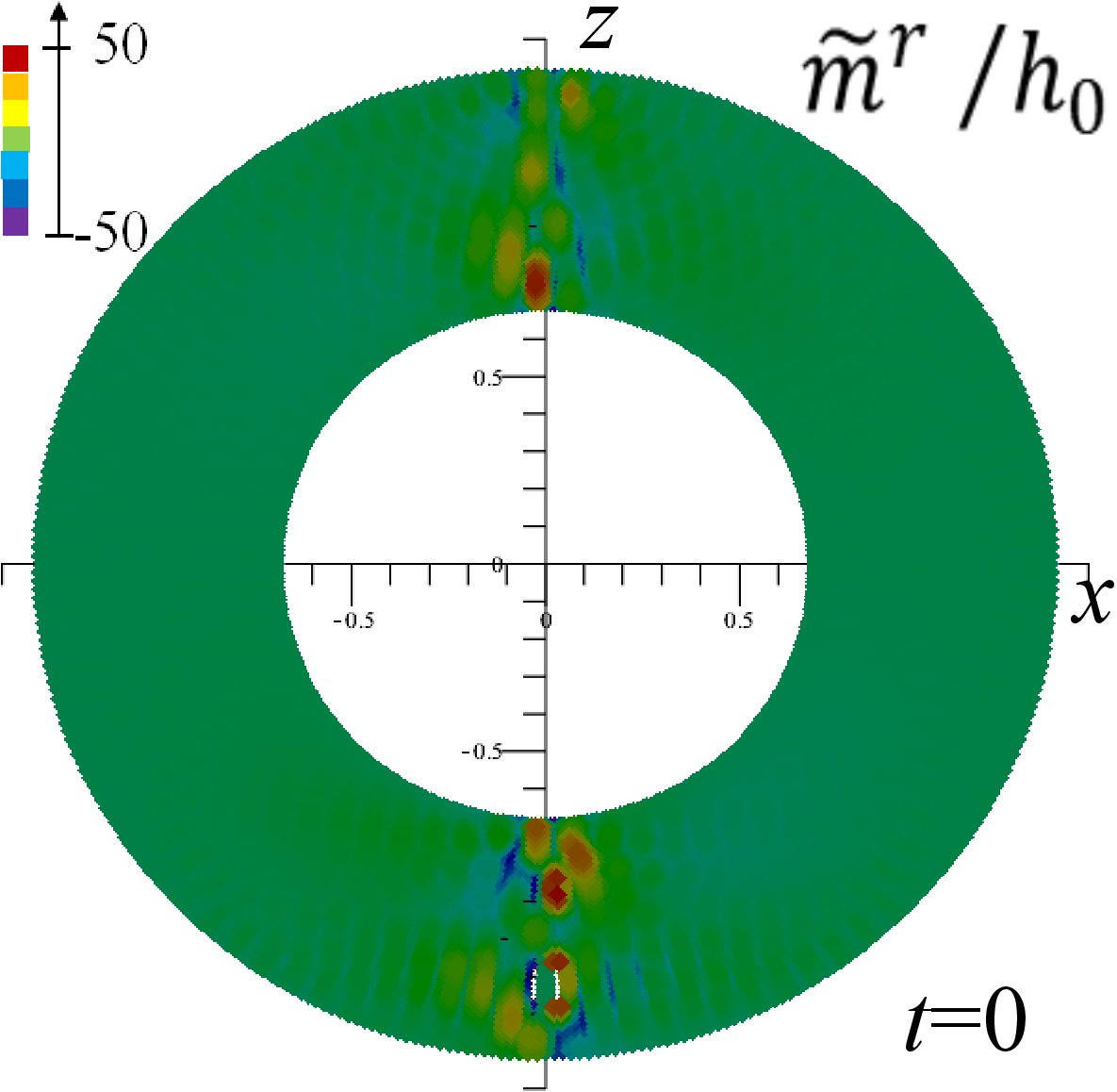}
\includegraphics[width=0.49\columnwidth]{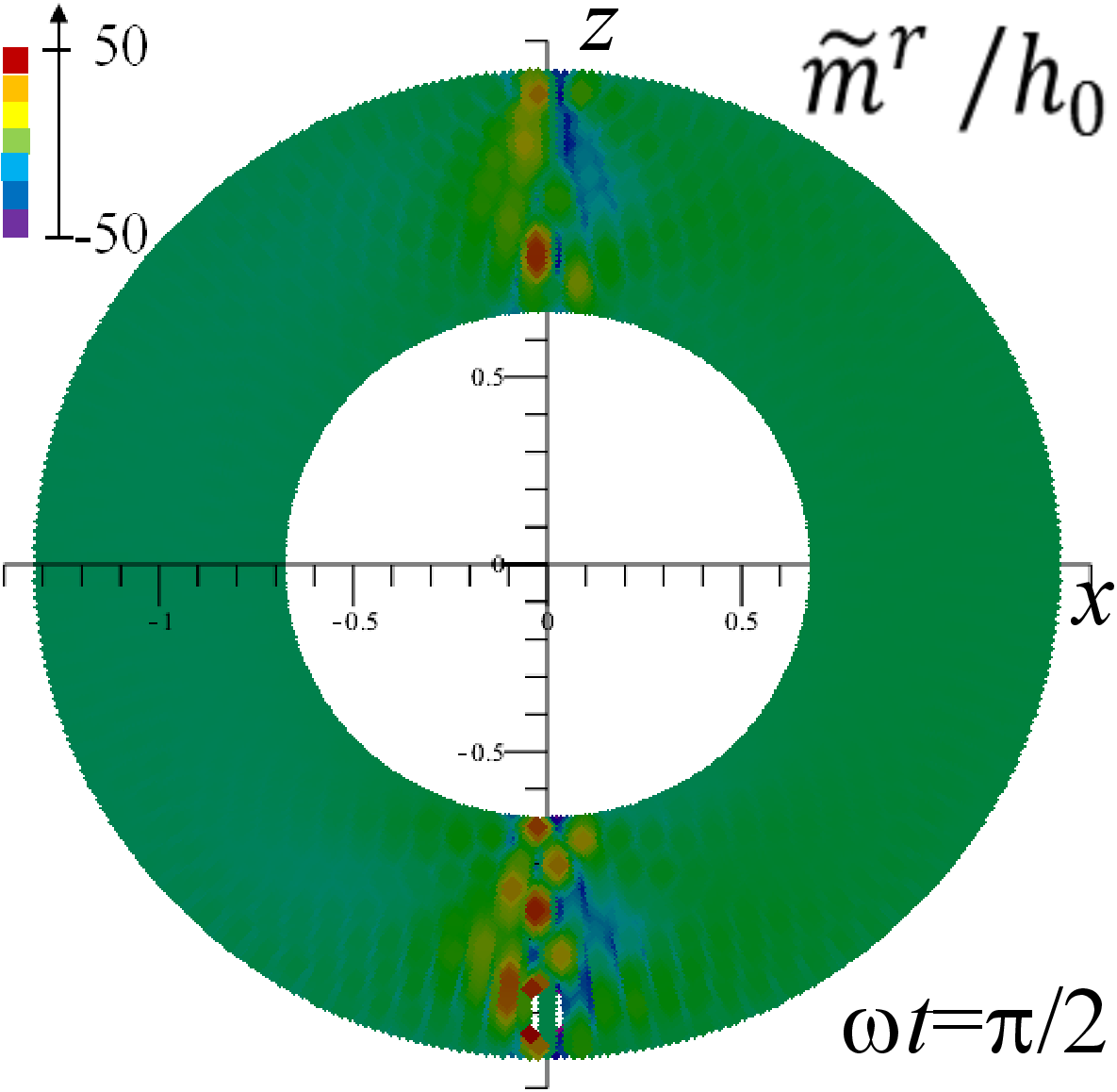}\vspace{5mm}\\
\includegraphics[width=0.49\columnwidth]{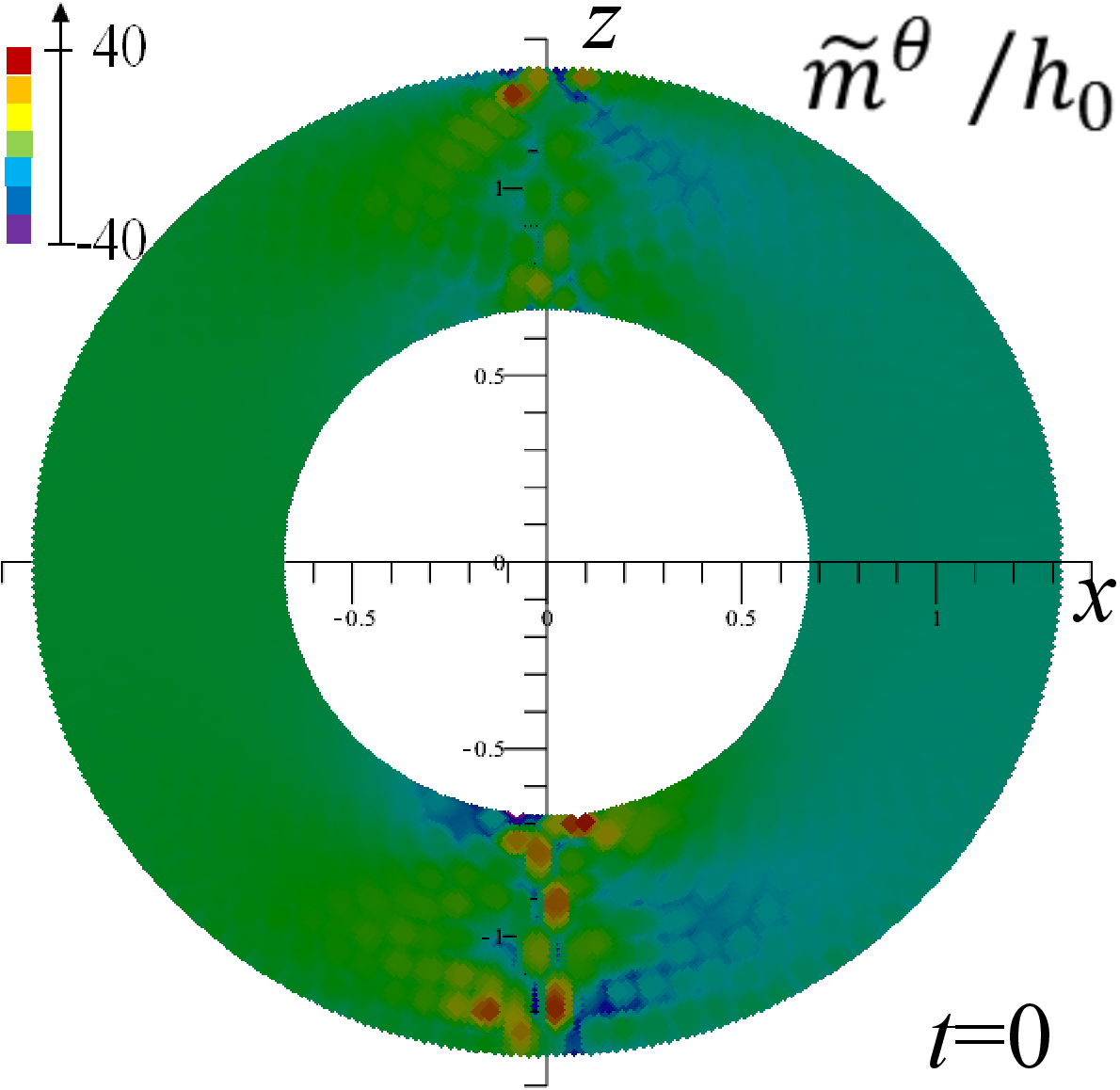}
\includegraphics[width=0.49\columnwidth]{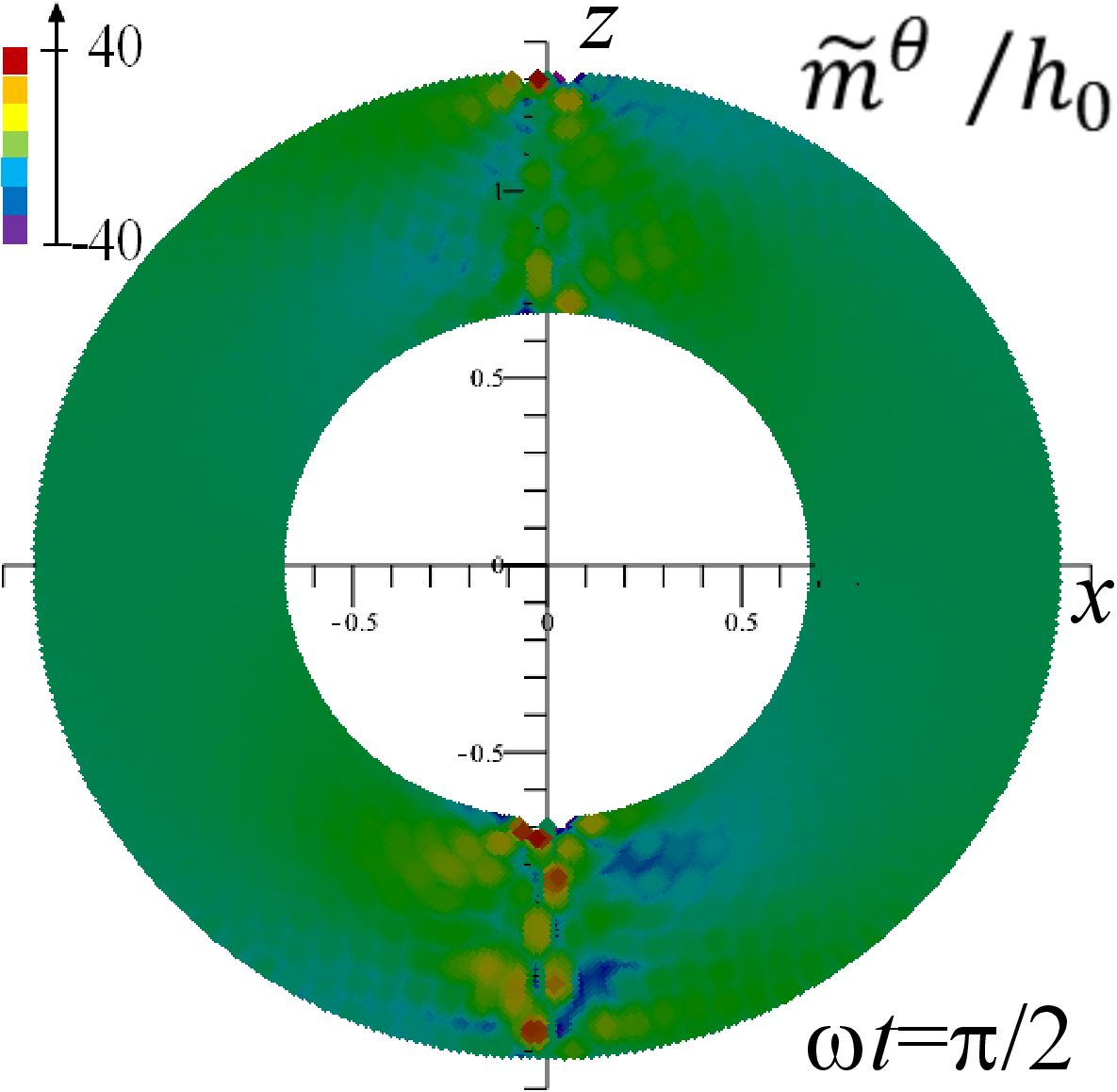}
\caption{\label{FigMX} The color plots of the radial (upper panels) and polar (lower panels) components of the particle's magnetization in the $xz$ plane, calculated at different $t$ in the range $0\leq\omega t\leq\pi$. The oscillation frequency corresponds to resonance marked by the arrow in Fig.~\ref{FigMuvsDelta}(b). External ac magnetic field is parallel to the $x$ axis. Model parameters: $\delta=0.3$, $K=0.1$ (type-II), $a_{ex}=0.01$, $\alpha=0.01$.}
\end{figure}

Let us consider now the case when the magnetic field is parallel to the $x$ axis. In this case the particle's magnetization is given by the formula
\begin{eqnarray}\label{mprec}
&&\tilde{m}^{a}(\mathbf{r},\,t)=-\Real\left\{\sum_{k=0}^{\infty}\sum_{l=1}^{\infty}f_k(r)\sqrt{\frac{2l+1}{6\pi l(l+1)}}P^{1}_{l}(\cos\theta)\times\right.\nonumber\\
&&\left.\left[\tilde{\chi}^{akl1}_{r011}(\omega)+\sum_{l'=1}^{\infty}\tilde{\chi}^{akl1}_{\theta0l'1}(\omega)D^{(1)}_{1l'}\right]e^{-i\omega t}\right\}\!\cos\varphi\,h_0\,.
\end{eqnarray}
We see that the magnetization is controlled now by the susceptibility matrix $\tilde{\chi}^{aklm}_{bk'l'm}(\omega)$ with $m=1$. The magnetization is proportional to $\cos\varphi$, that is, the largest amplitude of oscillations exists in the $xz$ plane. Our calculations show that at the main resonance frequency, the magnetic moments of the particle are excited only near the vortex core, and the amplitude of these oscillations is smaller than that for the case of the magnetic field parallel to the $z$ axis. Thus, the main contribution to the imaginary part of the permeability of the composite at the main resonance frequency comes from the particles with local $z$ axes parallel to the applied ac field.

Besides main resonance peak, the imaginary part of the permeability of the composite shows also several smaller peaks, laying both below and above main resonance. We study the magnetization distribution at different pronounced secondary resonances and found similar behavior of the magnetization for all considered frequencies. We show the results corresponding to the frequency marked by the arrow in Fig.~\ref{FigMuvsDelta}(b). Our calculations show that for all secondary resonances studied the largest amplitude of oscillations exists when the magnetic field is {\it perpendicular} to the $z$ axis of the particle. Figure~\ref{FigMX} shows the distribution of $\tilde{m}^{a}(\mathbf{r},\,t)$ in the $xz$ plane ($\cos\varphi=\pm1$) calculated at $\omega t=0$ and $\omega t=\pi/2$ (magnetic field is parallel to the $x$ axis). We see that the magnetization is excited only near the vortex core. The amplitude of these oscillations is several times larger than that at the main resonance. Thus, the main contribution to secondary resonances shown in Fig.~\ref{FigMuvsDelta} comes from the particles with local $z$ axes perpendicular to the ac magnetic field.

\begin{figure}[t]
\centering
\includegraphics[width=0.99\columnwidth]{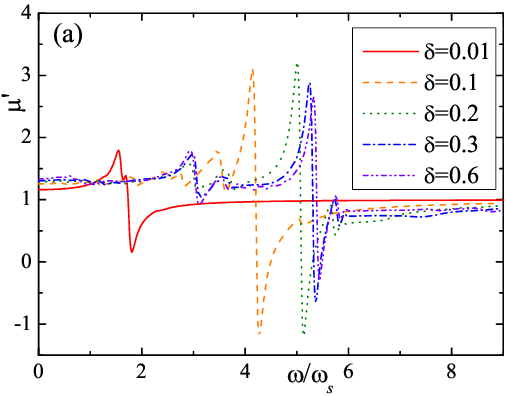}\\
\includegraphics[width=0.99\columnwidth]{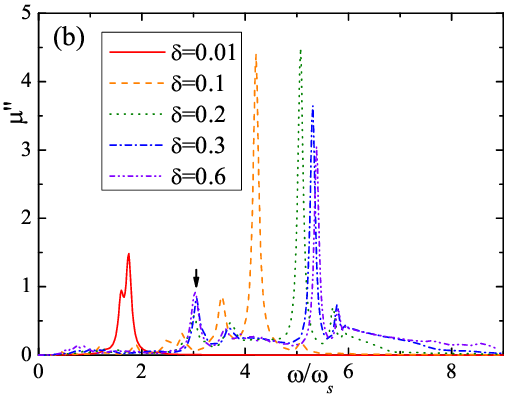}
\caption{\label{FigMuvsDeltaI} Frequency dependencies of real (a) and imaginary (b) parts of the permeability of the composite, calculated for five values (see the legend) of the shell thickness $d$. Model parameters: $K=-0.1$ (type-I), $a_{ex}=0.01$, $p_c=0.15$, $\alpha=0.01$.}
\end{figure}

Let us now discuss the results for the type-I magnetic anisotropy. Figure~\ref{FigMuvsDeltaI} shows the frequency dependencies of the real and imaginary parts of the composite's permeability calculated for different shell's thicknesses. The shell's thicknesses as well as other model parameters (with the exception of $K$) are the same as for Fig.~\eqref{FigMuvsDelta}. We see that with the exception of the case $\delta=0.01$ the curves $\mu$ vs $\omega$, shown in Fig.~\ref{FigMuvsDeltaI}, are practically the same as for the case of the type-II magnetic anisotropy, Fig.~\ref{FigMuvsDelta}. Analysis also shows that special profiles of the particle's magnetization are similar to that shown in Figs.~\ref{FigM1} and~\ref{FigMX}. Thus, for not too thin shells, the response of the system to the alternating magnetic field is similar for both types of the magnetic anisotropy considered. We attribute this to the fact that the leading terms in Landau-Lifshitz equation~\eqref{LLeqKLM} come from the ``demagnetizing factors'' $F^{ab}_{kk'l}$, while the role of the magnetic anisotropy is minor one. For the thin shell ($\delta=0.01$), the curve $\mu''$ vs $\omega$ shown in Fig.~\ref{FigMuvsDeltaI} has a double peak structure and is wider than that shown in Fig.~\ref{FigMuvsDelta}. This peak broadening will be explained in the next Section where we consider the limit $\delta\to0$.

\section{Permeability of the composite in the limit $d/R_0\to0$ and $l_{ex}/R_0\to0$}\label{PermeabilityLimit}

\begin{figure}[t]
\centering
\includegraphics[width=0.99\columnwidth]{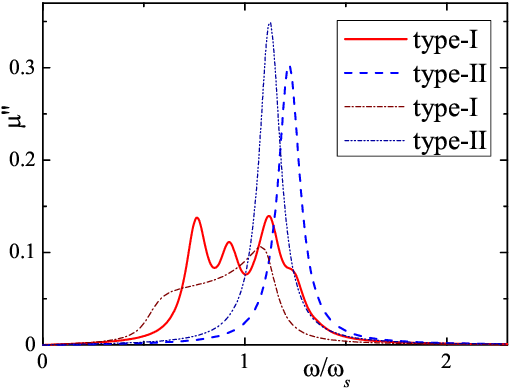}
\caption{\label{FigImMuI_II} Frequency dependencies of the imaginary part of the permeability of the composite with thin shells, calculated for two types of magnetic anisotropy (see the legend). Thick curves correspond to the numerical result according to Eq.~\eqref{LLeqKLM}, while thin curves correspond to the approximate formulas~\eqref{mucompas} and~\eqref{muq} (see the text below). Model parameters: $\delta=0.001$, $|K|=0.1$, $a_{ex}=0.01$, $p_c=0.15$, $\alpha=0.01$.}
\end{figure}

As it was shown above, the frequency dependencies of the composite's permeability are very similar to each other for two types of magnetic anisotropy when the thickness of the shell is not too small. However, at small $\delta$ the curves $\mu$ vs $\omega$ calculated for the type-I and type-II magnetic anisotropy (for the same $|K|$) are substantially different. For example, Fig.~\ref{FigImMuI_II} shows the imaginary parts of the composite's permeability calculated for two types of magnetic anisotropy for $\delta=0.001$. The curve for type-II anisotropy is just a single-peak, while the curve for the easy-plane anisotropy has a multi-peak structure, and the width of this (multi)peak is much wider than that for the type-II anisotropy.

In this section, we will explain this discrepancy and derive formulas for the permeability of the composite in the limit of $d/R_0\to0$ and $a_{ex}=l_{ex}/R_0\to0$ for two types of magnetic anisotropy considered. Let us start with the type-II magnetic anisotropy. For the thin shell, one can consider only first, homogeneous, radial mode, ignoring other modes. In the limit $a_{ex}\to 0$, we can also neglect the contributions to the Landau-Lifshitz equation~\eqref{LLeqKLM} from the energy of the exchange interaction. Next, according to the formula~\eqref{Fas}, we have $F^{r\theta}_{00l}\to0$, $F^{\theta\theta}_{00l}\to0$, and $F^{rr}_{00l}\to2l+1$ when $\delta\to0$. Ignoring the contributions from $F^{r\theta}_{00l}$ and $F^{\theta\theta}_{00l}$, we obtain that the equations~\eqref{LLeqKLM} become diagonal on the index $l$. Such equations can be solved analytically. Substituting the found expressions for $\tilde{\chi}^{a0lm}_{b0lm}$ into the formula~\eqref{mucomp}, we obtain the expression for the magnetic permeability of the composite in the form
\begin{eqnarray}
\mu(\omega)&=&1+\frac{p_F}{3}(4\pi\omega_s)^2\times\label{mucompas}\\
&&\frac{\left[A(1+K/4\pi)+K/4\pi\right]-i\alpha\omega(1+A)/(4\pi\omega_s)}
{\omega_r^2-(1+\alpha^2)\omega^2-2i\alpha\omega\omega_s(2\pi+K)},\nonumber
\end{eqnarray}
where
\begin{equation}\label{omegar}
\omega_r=\omega_s\sqrt{K(4\pi+K)}
\end{equation}
is the resonance frequency, and
\begin{equation}\label{A}
A=\frac13\left(\sum_{l=0}^{\infty}D^{(0)}_{1l}\right)^2+\frac23\left(\sum_{l=0}^{\infty}D^{(1)}_{1l}\right)^2.
\end{equation}
Performing numerical summation in the formula~\eqref{A}, we obtain $A\cong1.76$. Note that in the considered limit, the resonance frequency~\eqref{omegar} coincides with the resonance frequency of the thin film with uniaxial magnetic anisotropy. In addition, the expression~\eqref{mucompas} for the permeability is similar to that for the thin film (after averaging over the direction of the magnetic field), with the only difference that for the thin film $A\equiv1$.

\begin{figure}[t]
\centering
\includegraphics[width=0.99\columnwidth]{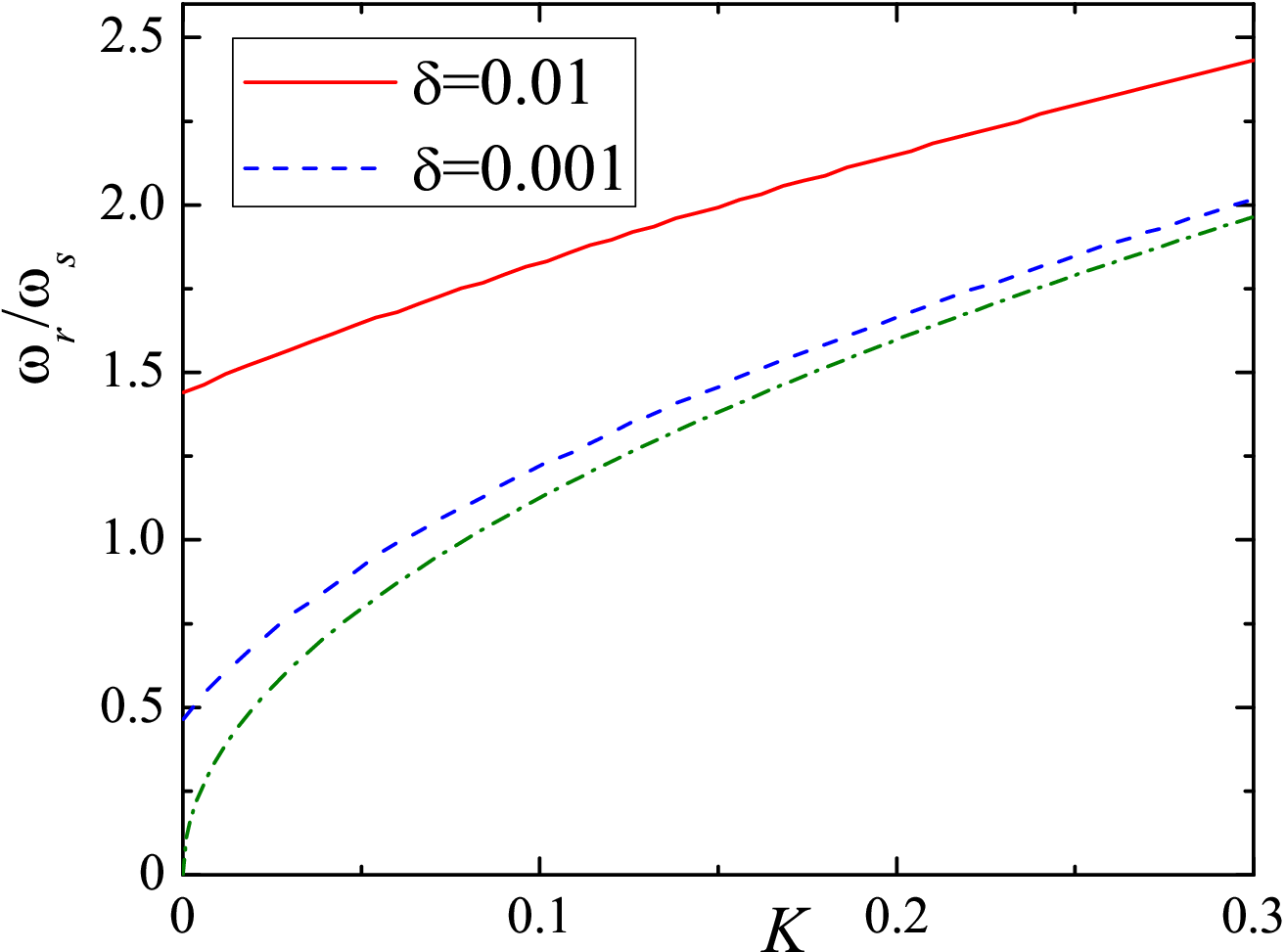}
\caption{\label{FigOmegavsBeta} The dependence of the main resonance frequency on the anisotropy parameter $K>0$ (type-II), calculated for the shell thickness $\delta=0.01$ (red solid curve) and $\delta=0.001$ (blue dashed curve). Green dot-dashed curve corresponds to Eq.~\eqref{omegar}. Model parameters: $a_{ex}=0.01$, $p_c=0.15$, $\alpha=0.01$.}
\end{figure}

Figure~\ref{FigOmegavsBeta} shows the dependence of the main resonance frequency on the anisotropy parameter $K$, calculated numerically using Eq.~\eqref{LLeqKLM} for two different shell thicknesses $\delta$ (in these calculations we take $K_{\text{max}}=10$). The curve $\omega_r(K)$ with $\omega_r$ from Eq.~\eqref{omegar} is also shown in this figure for comparison. We see, that Eq.~\eqref{omegar} underestimate the resonance frequency. For very thin shell ($\delta=0.001$) formula~\eqref{omegar} is adequate approximation for the resonance frequency when the anisotropy parameter is not very close to zero. However, already for $\delta=0.01$, disagreement between analytical expression~\eqref{omegar} and the numerical result is quite strong. This can be explained by the fact that for such $\delta$ we cannot neglect the ``demagnetizing factors'' $4\pi F^{\theta\theta}_{00l}\cong8\pi\delta$ and $4\pi F^{r\theta}_{00l}\cong-4\pi\delta$ in Eq.~\eqref{LLeqKLM} in comparison to $K$.

Analysis shows that formula~\eqref{mucompas} overestimates the amplitude of the resonance peak. This is explained as follows. In formula~\eqref{A}, the summation over $l$ is performed up to infinity. At the same time, we can neglect the contributions from $F^{r\theta}_{00l}$ and $F^{\theta\theta}_{00l}$ only if $l\delta\ll1$. In addition, deriving formula~\eqref{mucompas}, we neglected the contributions from the exchange energy, which can be done only when $la_{ex}\ll1$. The latter inequality fails when $l\gtrsim10$. Limiting the summation in formula~\eqref{A} to $l=L_0=10$, we obtain $A(L_0=10)\cong1.05$. The corresponding curve of the imaginary part of the permeability is shown in Fig.~\eqref{FigImMuI_II} by the thin dashed-dot-dot curve. We see a good agreement between the approximate formula~\eqref{mucompas} and the numerical result.

Let us consider now the type-I magnetic anisotropy. In the approximations described above the Landau--Lifshitz equations~\eqref{LLeqKLM} take the form
\begin{eqnarray}
&&-\frac{1}{\omega_s}\frac{\partial\tilde{m}^{\theta lm}}{\partial t}+\frac{\alpha}{\omega_s}\frac{\partial\tilde{m}^{rlm}}{\partial t}
+\left(4\pi-\frac{K}{2}\right)\tilde{m}^{rlm}\nonumber\\
&&-K\sum_{l'=|m|}^{\infty}\left[C_{ll'}^{(m)}\tilde{m}^{rlm}-S_{ll'}^{(m)}\tilde{m}^{\theta lm}\right]=\tilde{h}^{rlm}\,,\nonumber\\
&&\frac{1}{\omega_s}\frac{\partial\tilde{m}^{rlm}}{\partial t}+\frac{\alpha}{\omega_s}\frac{\partial\tilde{m}^{\theta lm}}{\partial t}
-\frac{K}{2}\tilde{m}^{\theta lm}\nonumber\\
&&+K\sum_{l'=|m|}^{\infty}\left[C_{ll'}^{(m)}\tilde{m}^{\theta lm}+S_{ll'}^{(m)}\tilde{m}^{rlm}\right]=\tilde{h}^{\theta lm}\,,\label{LLeqLM}
\end{eqnarray}
where $\tilde{m}^{alm}=\tilde{m}^{a0lm}$ and $\tilde{h}^{alm}=\tilde{h}^{a0lm}$. Analysis shows that diagonal and non-diagonal (in indices $l$ and $l'$) elements of the coefficients $C_{ll'}^{(m)}$ and $S_{ll'}^{(m)}$ are of the same order of magnitude. Thus, in contrast to the case of the type-II magnetic anisotropy, equations~\eqref{LLeqLM} cannot be decoupled. This system can be solved only numerically. However, one can derive an approximate analytical formula for the permeability if we consider the limit $l,\,l'\to\infty$. In this case one can derive analytical formulas for coefficients $C_{ll'}^{(m)}$ and $S_{ll'}^{(m)}$. Using expressions~\cite{hobson1931theoryYlm} for the asymptotic of the spherical harmonics at large $l$, we obtain
\begin{eqnarray}
C_{ll'}^{(m)}&\cong&\frac12\left[\delta_{l,l'-2}+\delta_{l,\,l'+2}\right]\,,\label{CSas}\\
S_{ll'}^{(m)}&\cong&-\!\!\sum_{n=-\infty}^{\infty}\!\!t_n\,\delta_{l-l',\,2n-1}\,,\;
t_n=\frac{4}{\pi}\frac{1}{(2n+1)(2n-3)}\,.\nonumber
\end{eqnarray}
We see that in the considered limit, the $C_{ll'}^{(m)}$ and $S_{ll'}^{(m)}$ do not depend on $m$ and depend on the difference $l-l'$. The latter circumstance allows us to seek a solution to Eq.~\eqref{LLeqLM} in the form $\tilde{m}^{alm}=\tilde{m}^{am}_{q}e^{iql}$, where the ``wave vector'' $q$ varies in the range $0<q<2\pi$. Using Eq.~\eqref{CSas} and performing the Fourier transform in Eq.~\eqref{LLeqLM}, we obtain the equations for $\tilde{m}^{am}_{q}$, which we write in the matrix form as
\begin{eqnarray}
&&\left(\begin{array}{cc}
4\pi-K\cos^2\!q-\frac{i\omega\alpha}{\omega_s}&\frac{i\omega}{\omega_s}+\frac{K}{2}s_q\sin2q\\
-\frac{i\omega}{\omega_s}+\frac{K}{2}s_q\sin2q&-K\sin^2\!q-\frac{i\omega\alpha}{\omega_s}
\end{array}\right)
\left(\begin{array}{c}
\tilde{m}^{rm}_{q}\\
\tilde{m}^{\theta m}_{q}
\end{array}\right)=\nonumber\\
&&=\left(\begin{array}{c}
\tilde{h}^{rm}_{q}\\
\tilde{h}^{\theta m}_{q}
\end{array}\right),\label{LLeqQ}
\end{eqnarray}
where $s_q=1$ if $0<q<\pi$ and $s_q=-1$ if $\pi<q<2\pi$. Deriving this equation, we used the following equality~\cite{prudnikov1988integrals}
\begin{equation}
\sum_{n=-\infty}^{\infty}t_ne^{-i(2n-1)q}=-s_q\sin2q\,.
\end{equation}
Equating the determinant of the matrix in Eq.~\eqref{LLeqQ} to zero (at $\alpha=0$) we obtain the resonance frequency at given $q$. It is equal to (remember that $K<0$ in the case under study)
\begin{equation}\label{omegaq}
\omega_q=\omega_s\sqrt{-4\pi K}|\sin q|\,.
\end{equation}
Thus, instead of single resonance frequency, Eq.~\eqref{omegar}, characteristic to the type-II magnetic anisotropy, now we have a band of resonance frequencies laying in the range $0<\omega_q<\sqrt{-4\pi K}$. This explains the peak broadening observed for thin shells (see Figs.~\ref{FigMuvsDelta}, \ref{FigMuvsDeltaI}, and~\ref{FigImMuI_II}).

Solving Eq.~\eqref{LLeqQ}, we find $\tilde{\chi}^{a0lm}_{b0l'm}$, and using the formula~\eqref{mucomp}, we finally obtain the formula for the permeability of the composite in the form
\begin{eqnarray}
\mu(\omega)&=&1+\frac{4\pi p_F\omega_s^2}{3}\int\limits_0^{\pi}\!\frac{dq}{\pi}\times\label{muq}\\
&&\frac{-K\sin^2q-\frac{i\omega\alpha}{\omega_s}+(4\pi-K\cos^2q-\frac{i\omega\alpha}{\omega_s}){\cal D}_q}
{\omega_q^2-(1+\alpha^2)\omega^2-i\omega\omega_s\alpha(4\pi-K)}\,,\nonumber
\end{eqnarray}
where
\begin{equation}\label{Dq}
{\cal D}_q=\frac13\left|\sum_{l=0}^{\infty}e^{iql}D_{1l}^{(0)}\right|^2+\frac23\left|\sum_{l=1}^{\infty}e^{iql}D_{1l}^{(1)}\right|^2\,.
\end{equation}
Analysis shows that formula~\eqref{muq} correctly reproduces the high-frequency edge of the permeability peak. However, it fails at low frequencies. For example $\mu''(\omega)$ diverges when $\omega\to0$ since $\omega_q\to0$ when $q\to0$ or $\pi$. We have to remember, however, that formula~\eqref{CSas} works only at large $l$ and $l'$: for small orbital indices the coefficients $C_{ll'}^{(m)}$ and $S_{ll'}^{(m)}$ are substantially different from their asymptotics. As a result, the band of the resonance frequencies becomes restricted from below. This fact can be taken into account by the restriction of the $q$ integration in Eq.~\eqref{muq} by the range $q_0<q<\pi-q_0$, where $q_0$ is the fitting parameter. The dependence $\mu''(\omega)$ calculated at $q_0=0.5$ is shown Fig.~\ref{FigImMuI_II} by thin dot-dashed curve. We see that the approximate formula~\eqref{muq} underestimate the amplitude of the peak, and it cannot explain the multi-peak structure observed from the numeric calculations.

\section{Stability of the static magnetic configuration. Self-oscillations of the magnetization}\label{Stability}

\begin{figure}[t]
\centering
\includegraphics[width=0.99\columnwidth]{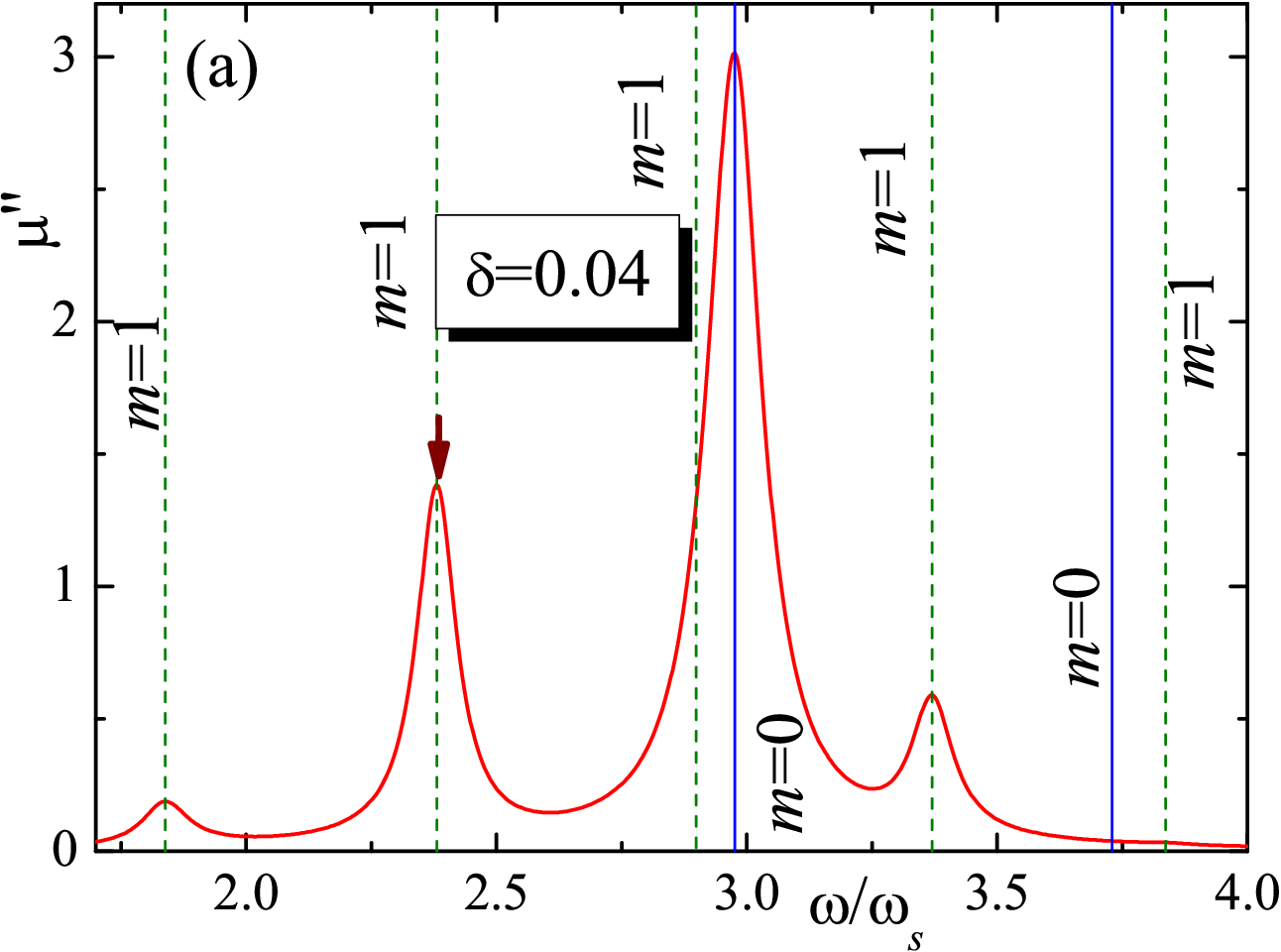}
\includegraphics[width=0.99\columnwidth]{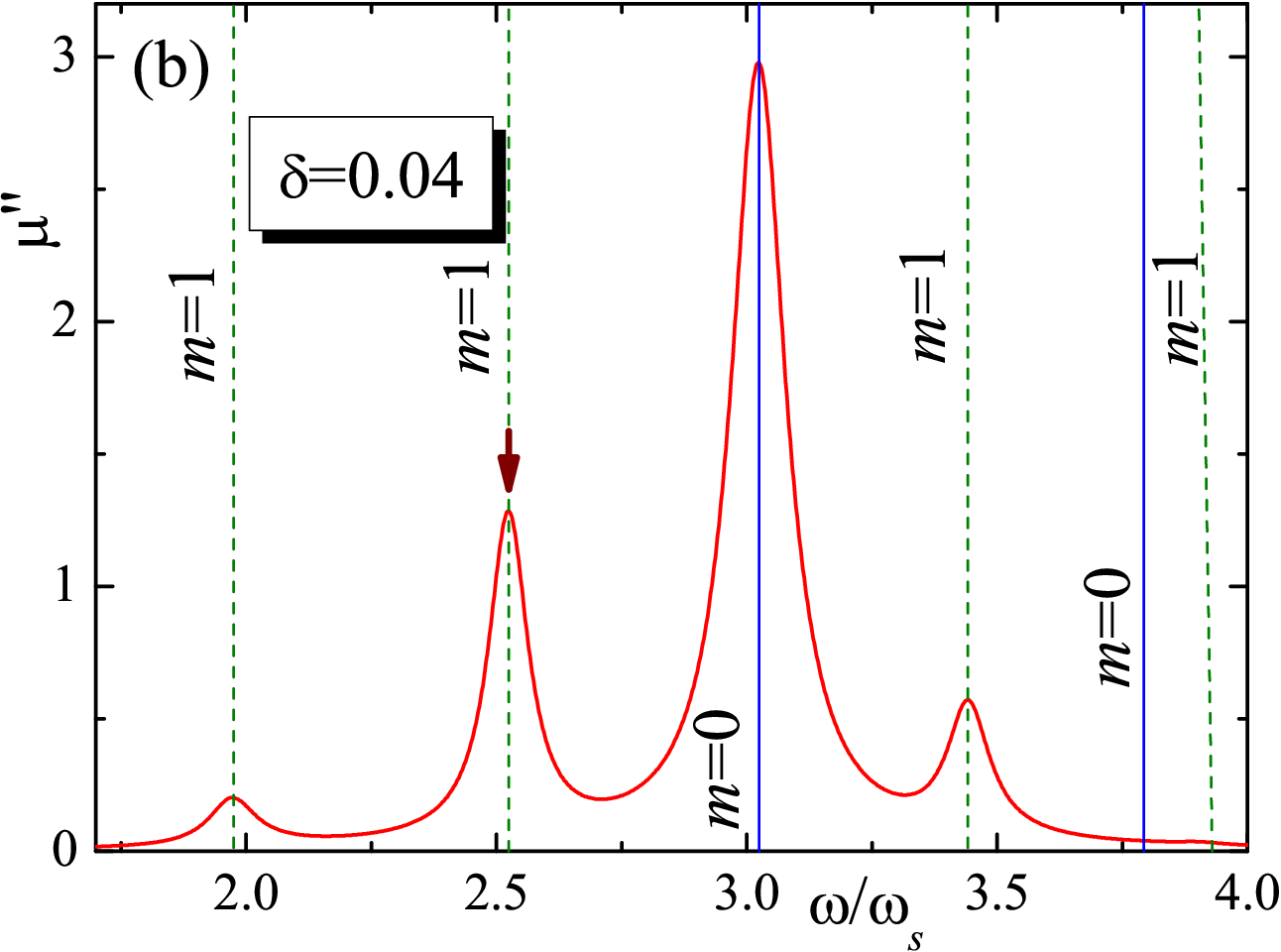}
\caption{\label{FigImMuD004} The frequency dependence of the imaginary part of the permeability calculated for the type-I (a) and type-II (b) magnetic anisotropy. Parameters of the model are: $\delta=0.04$, $|K|=0.1$, $a_{ex}=0.01$, $p_c=0.15$, $\alpha=0.01$. Vertical lines mark the square roots of the eigenvalues of the matrices $\hat{\Omega}^2_0$ (solid lines) and $\hat{\Omega}^2_1$ (dashed lines).}
\end{figure}

In this section we study the self-oscillations of the magnetization and show that the vortex magnetic configuration shown in Fig.~\ref{FigConfig} more likely corresponds to the minimum of energy. In order to obtain the equations for self-oscillations, it is necessary to take in Eq.~\eqref{LLeqMatrix} $\alpha=0$ and $\tilde{h}^{a}_m=0$. As a result, Eq.~\eqref{LLeqMatrix} can be rewritten as
\begin{equation}\label{EigenEq}
\omega^2
\left(\begin{array}{c}
\tilde{m}_{m}^{r}\\ \tilde{m}_{m}^{\theta}
\end{array}\right)
+\hat{\Omega}_m^2
\left(\begin{array}{c}
\tilde{m}_{m}^{r}\\ \tilde{m}_{m}^{\theta}
\end{array}\right)=0\,,
\end{equation}
where
\begin{equation}\label{Omega2}
\hat{\Omega}_m^2=\left(\begin{array}{cc}
\hat{\Lambda}_m^{\theta\theta}\hat{\Lambda}_m^{rr}-\left(\hat{\Lambda}_m^{\theta r}\right)^2&
\hat{\Lambda}_m^{\theta\theta}\left(\hat{\Lambda}_m^{\theta r}\right)^T\!\!-\hat{\Lambda}_m^{\theta r}\hat{\Lambda}_m^{\theta\theta}\\
\hat{\Lambda}_m^{rr}\hat{\Lambda}_m^{\theta r}-\left(\hat{\Lambda}_m^{\theta r}\right)^T\!\!\hat{\Lambda}_m^{rr}&
\hat{\Lambda}_m^{rr}\hat{\Lambda}_m^{\theta\theta}-\left[\left(\hat{\Lambda}_m^{\theta r}\right)^T\right]^2
\end{array}\right).
\end{equation}
This equation is the eigenvalue equation for the matrix $\hat{\Omega}_m^2$. We are only interested in matrices with $m=0$ and $m=\pm1$, since only oscillations with such $m$ are excited in the homogeneous magnetic field. It can also be shown that the matrices with $m=+1$ and $m=-1$ are identical. Therefore, we only examine the matrices $\hat{\Omega}_0^2$ and $\hat{\Omega}_1^2$. The analysis shows that for all values of the model parameters considered in this paper, the eigenvalues of these matrices, $\omega^2_{mS}$ ($S=1,\,2,\,\dots,\,2K_{\max}L_{\max}$), turn out to be real and positive. It indicates~\cite{Note3} that the equilibrium magnetic configuration considered here indeed corresponds to the minimum of energy (of course, not necessarily global). In addition, it turns out that all eigenvalues of the matrices $\hat{\Omega}_m^2$ are doubly degenerate.

For large $K_{\text{max}}$ and $L_{\text{max}}$ the set of eigenfrequencies $\omega_{mS}$ is quite dense. For this reason, to be more illustrative, we consider here the thin shell taking into account only one, homogeneous, radial mode. Figure~\ref{FigImMuD004} shows the frequency dependence of the imaginary part of the composite's permeability, calculated at $\delta=0.04$ for both types of magnetic anisotropy. The curves $\mu''$ vs $\omega$ are similar to each other, both of them have multi-peak structure. Thus, the case $\delta=0.04$ cannot be considered as belonging to the $\delta\to0$ limit described in the previous Section. The vertical lines in Figs.~\ref{FigImMuD004}(a, b) show all $\omega_{0S}$ and $\omega_{1S}$ laying within the frequency range $1.7<\omega/\omega_s<4$. We see that the main resonance corresponds to some eigenfrequency $\omega_{0S}$, while all side resonances correspond to the eigenfrequencies $\omega_{1S}$. Note also that there are some eigenfrequencies that do not correspond to any peaks in the imaginary part of the permeability. This can be explained by the presence of a selection rule prohibiting the excitation of certain self-oscillations in a homogeneous field. These selection rules are apparently controlled by the parameters $D^{(m)}_{1l}$ [see formula~\eqref{mucomp}], for which the equality $D^{(m)}_{1l}=0$ is valid for $\mod(l,\,2)=1$ due to the symmetry of spherical harmonics.

\begin{figure}[t]
\centering
\includegraphics[width=0.49\columnwidth]{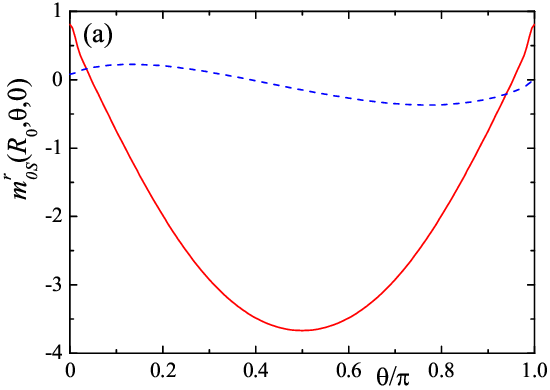}
\includegraphics[width=0.49\columnwidth]{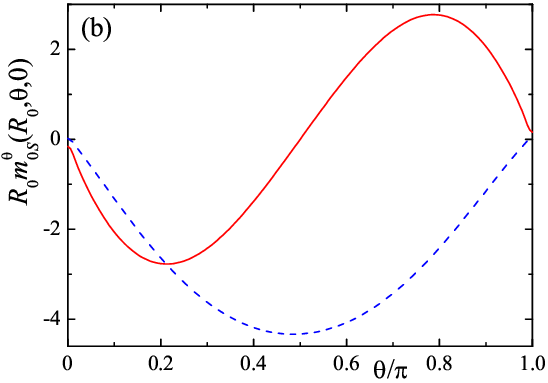}\\
\includegraphics[width=0.49\columnwidth]{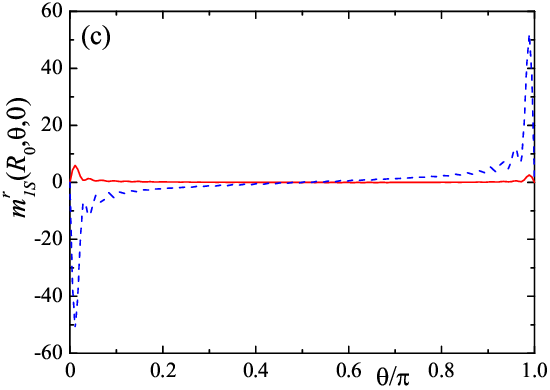}
\includegraphics[width=0.49\columnwidth]{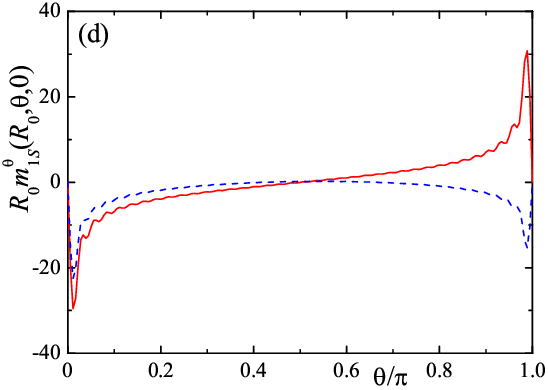}
\caption{\label{FigMA} The dependencies of the radial (a, c) and polar (b, d) components of the self magnetization oscillations on the polar angle $\theta$ corresponding to the main resonance (panels a and b) and the resonance indicated by the arrow in Fig.~\ref{FigImMuD004} (panels c and d). Model parameters: $\delta=0.04$, $K=-0.1$ (type-I), $a_{ex}=0.01$.}
\end{figure}

Let us now consider the spatial profiles of the self-oscillations of the magnetization corresponding to some resonant frequencies. If $\tilde{m}_{S}^{aklm}$ is some eigenfunction of the matrix $\hat{\Omega}_{m}^2$, then the magnetization fluctuations, corresponding to this mode, are calculated by the formula
\begin{equation}\label{mMS}
\tilde{m}^{a}_{Sm}(\mathbf{r})=\sum_{k=0}^{\infty}\sum_{l=|m|}^{\infty}\tilde{m}_{S}^{aklm}f_k(r)Y_{l}^{m}(\theta,\,\varphi)\,.
\end{equation}
When only one, homogeneous, radial mode is taken into account, the functions $\tilde{m}^{a}_{Sm}(\mathbf{r})$ do not depend on $r$. Further, for $m=0$ these functions do also not depend on $\varphi$, while for $m=\pm1$ this dependence reduces to $e^{\pm i\varphi}$. Since all eigenfrequencies turn out to be doubly degenerate, each eigenfrequency will correspond to two eigenfunctions $\tilde{m}_{S}^{aklm}$. Calculations show that all eigenfunctions of $\tilde{m}_{S}^{aklm}$ are real-valued. Figures~\ref{FigMA}(a,b) show the dependencies $\tilde{m}^{r}_{S0}(\mathbf{r})$ and $\tilde{m}^{\theta}_{S0}(\mathbf{r})$ on the polar angle $\theta$ for the eigenfunctions of the matrix $\hat{\Omega}^2_0$ corresponding to the main resonance frequency. We see that the oscillations at the poles are practically not excited. In this case, the profiles $\tilde{m}^{a}_{S0}(\mathbf{r})$ are in many ways similar to those for the magnetization fluctuations in the ac field parallel to the $z$ axis of the particle at times $\omega t=0$ and $\omega t=\pi/2$ (see Fig.~\ref{FigM1}). We observe a completely different picture for the self oscillations corresponding to the matrix $\hat{\Omega}^2_1$. Figures~\ref{FigMA}(c, d) show the dependencies of $\tilde{m}^{a}_{S1}(R_0,\,\theta,\,0)$ on the angle $\theta$ for the eigenfunctions corresponding to the resonance frequency indicated by the arrow in the Fig.~\ref{FigImMuD004}. We see that fluctuations occur mainly near the poles of the particle. This correlates with the results shown in Fig.~\ref{FigMX}. The results in Fig.~\ref{FigMA} are shown for the type-I magnetic anisotropy. Similar results are obtained for the type-II magnetic anisotropy.

\section{Discussion and conlcusions}\label{Discussion}

Thus, we studied the response of the composite consisting of hollow spherical ferromagnetic particles to an alternating magnetic field. We analyzed the behavior of the magnetic permeability as a function of the ratio $\delta=2d/R_0$. We showed that the main resonance frequency shifts to the right as $\delta$ increases. This behavior seems natural and finds its analogy with the behavior of resonance frequency in a thin film. Indeed, for a film with longitudinal dimensions $L$ and thickness $d$, the components of the demagnetization tensor along the film, $N_{x,y}$, increase with the increase of the ratio $d/L$: $N_{x,y}\propto d/L\ln(L/d)$. In our case, the role of demagnetizing factors $N_{x,y}$ is played by the parameters $F^{r\theta}_{00l}$ and $F^{\theta\theta}_{00l}$, which increase linearly with $\delta$ at small $\delta$. This leads to a shift of the resonant frequency to the right. In the limit of $\delta\to0$ we obtained an analytical expressions for the permeability of the composite. For the type-II magnetic anisotropy the permeability is in many ways similar to that for a thin film.

The main assumption of this work is the vortex structure of the equilibrium magnetization configuration, schematically shown in Fig.~\ref{FigConfig} and described by Eq.~\eqref{M0} (outside the vortex core). Such kind of vortex magnetic configuration has been studied theoretically, in Refs.~\onlinecite{VortexConfiguration,VortexConfiguration2,VortexConfigurationExp}. Micromagnetic simulations were performed in Refs.~\onlinecite{VortexConfiguration,VortexConfigurationExp}. Thus, in Ref.~\onlinecite{VortexConfigurationExp} the calculated vortex configuration is very similar to our case (see Fig.~2 of that paper). In Ref.~\onlinecite{VortexConfiguration2} the authors calculated the equilibrium magnetic configuration minimizing the sum of the exchange and the easy-plane anisotropy energy. The function they calculated is $M^{a}(\theta)$. They found that outside the vortex core $M^{r}=0$. They also obtained that $M^{\theta}$ is non zero, but is much smaller than $M^{\varphi}$. However, they completely ignored the magnetostatic energy. At the same time, we have $\Div\mathbf{M}=g^{-1/2}\partial(\sqrt{g}M^{a})/\partial x^a=(1/\sin\theta)\partial(\sin\theta M^{\theta})/\partial\theta\neq0$. Thus, the magnetostatic energy is non-zero and tends to decrease $M^{\theta}$ making the ``in-surface'' vortex more favorable. There is also experimental evidence of the vortex magnetic configuration in spherical shells. Thus, the authors of Ref.~\onlinecite{VortexConfigurationExp} showed experimentally using the electron holography technique, that the static magnetic configuration of the hollow ferromagnetic particle indeed has the vortex-like structure.

The structure we consider is characterized by the presence of two vortex cores located at the poles of the particle, the influence of which we neglect. At the same time, we have shown that the side resonances of the permeability of the composite correspond to the magnetization oscillations  near the poles. Therefore, the question of whether these side resonances are an artifact of the approximation under consideration or whether they are really characteristic of the equilibrium magnetic configuration under consideration requires separate consideration. In particular, it is possible to perform a micromagnetic calculation of the equilibrium configuration and then solve the Landau-Lifshitz equations linearized around the found magnetization distribution.

\begin{comment}
Another significant assumption of this work is a special type of uniaxial magnetic anisotropy: the axis of anisotropy is not fixed, but rotates in space along with the magnetization. Strictly speaking, this leads to the fact that the magnetic configuration under consideration corresponds to a certain extreme of energy (a local minimum, as shown in Section~\ref{Stability}). If we assume that the particle has a certain fixed axis of anisotropy, this will generally lead to a change in its ground state magnetic configuration, which can significantly complicate calculations. Indeed, in this case we can expect more complicated dependence of the magnetization on the azimuthal angle $\varphi$. At the same time, in the case of weak magnetic anisotropy, $K\ll1$, these effects can be assumed to be insignificant.
\end{comment}

In this paper we completely neglected the skin effect. This can be done if the thickness $d$ of the shell is much smaller than the skin depth $l_s$. Let us evaluate here the characteristic thicknesses $d$ for which this can be done. The skin depth can be estimated using the formula~\cite{LandauVIIIen}
$$
l_s=\frac{c}{\sqrt{2\pi\sigma\omega\mu_{st}}}\,,
$$
where $c$ is the speed of light, while $\sigma$ and $\mu_{st}$ are the conductivity and static permeability of the particle, respectively. Let us estimate $l_s$ at the frequency corresponding to the main resonance for the composite with $\delta=0.2$, $\omega\cong5\omega_s$ (see Fig.~\ref{FigMuvsDelta}). Assuming $M_s=1700$\,Gs and $\sigma=9\times10^{16}$\,sec$^{-1}$ (as for Fe), we obtain $l_s\approx1/\sqrt{\mu_{st}}$\,$\mu$m. The static permeability of the particle can be estimated from the results obtained above using the formula $\mu_{st}=1+(\mu(0)-1)/p_F$, where $\mu(0)$ is the permeability of the composite at zero frequency. Thus, for $\delta=0.2$, we obtain $\mu_{st}\approx3.8$, and $l_s\approx0.5$\,$\mu$m. For particles' diameters $D\sim1$\,$\mu$m and $d\lesssim R_2$ we have $d\lesssim l_s$. Thus, we are at the edge of the applicability of the absence skin effect approximation. How the skin effect affects the results obtained will be considered in future studies.

In this paper we consider two types of magnetic anisotropy: easy-plane magnetic anisotropy (type-I) and ``circular'' uniaxial magnetic anisotropy (type-II). The latter is characterized by the rotation of the easy axis in the $xy$ plane. For both types of the magnetic anisotropy the static magnetic configuration we consider has a vortex-like structure, and is described by the same equation~\eqref{M0}. We showed that this magnetic configuration corresponds to the minimum of energy. The results obtained show similar behavior of the permeability and the self-oscillation modes for both types of anisotropy, when the thickness of the shell is not very small. This is explained by the fact that for not very thin shells and not too hard magnetic materials ($K\lesssim0.1$) the main terms in the derived Landau-Lifshitz equation~\eqref{LLeqKLM} come from the ``demagnetizing factors'' $F^{ab}_{kk'l}$ controlled by the magnetostatic energy. This opens the way to extend the proposed approach to any types of magnetic anisotropy. Indeed, for not very hard magnetic materials ($K<1$) one can consider the magnetic anisotropy energy by perturbations. The static magnetic configuration can be written as $\mathbf{M}=\mathbf{M}_0+\delta\mathbf{M}$, where $\mathbf{M}_0$ is given by Eq.~\eqref{M0}, while the correction $\delta\mathbf{M}$ is found by perturbations on the magnetic anisotropy constant using Eq.~\eqref{LL0}. When the equilibrium magnetic configuration is found the high-frequency permeability of the shell can be calculated by the approach described in this paper.

In conclusion, we studied the response to the external alternating magnetic field of the composite consisting of ferromagnetic spherical shells. We consider two types of magnetic anisotropy helping to stabilize the vortex-like structure of the static magnetic configuration of the shell. We showed that for not too thin shells and not very hard magnetic materials the high-frequency properties of the composite are similar for both type of magnetic anisotropy considered. We showed also that the considered static magnetic configuration corresponds to the minimum of energy. For the thin shells we obtained an approximate analytical formulas for the composite's permeability. The proposed approach can be directly extended for any type of the magnetic anisotropy when the magnetic material forming the shells is not hard.

\section*{Acknowledgments}

This work was supported by the Russian Science Foundation under grant No. 25-19-00393, https://rscf.ru/en/project/25-19-00393/. The author is grateful to A.\,N.~Lagarkov, K.\,N.~Rozanov, A.\,L.~Rakhmanov, and N.\,A.~Buznikov for useful discussions.

\appendix

\section{Construction of radial orthogonal functions}

In this Appendix, we describe an algorithm for construction of the radial functions orthogonal with weight $r^2$ and satisfying the boundary conditions~\eqref{LLboundF}. To do this, we will first represent them as
\begin{equation}\label{FtoG}
f_k(r)=\sqrt{\frac{V}{d}}\frac{g_k(x(r))}{r}\,,\;\;x(r)=\frac{r-R_1}{d}\,,\;\;0\leq x\leq1\,.
\end{equation}
The functions $g_k(x)$ should be orthogonal with weight $1$,
\begin{equation}\label{gknorm}
\int\limits_0^1dx\,g_k(x)g_{k'}(x)=\delta_{kk'}\,,
\end{equation}
and satisfy the boundary conditions
\begin{equation}\label{gkcond}
\left\{\begin{array}{l}
g'_k(0)-\displaystyle\frac{2\delta}{1-\delta}g_k(0)=0\\
g'_k(1)-\displaystyle\frac{2\delta}{1+\delta}g_k(1)=0
\end{array}
\right.\,,
\end{equation}
following from the conditions~\eqref{LLboundF}. The functions $g_k(x)$ can be constructed in different ways. Here we choose the following. Consider trigonometric functions of the form
\begin{equation}\label{yk}
y_k(x)=A_k\sin(\mu_kx)+B_k\cos(\mu_kx)\,.
\end{equation}
We will require them to satisfy the boundary conditions~\eqref{gkcond}. Then we obtain,
\begin{equation}\label{yk1}
y_k(x)=C_k\left[\frac{2\delta}{(1-\delta)\mu_k}\sin(\mu_kx)+\cos(\mu_kx)\right]\,,
\end{equation}
where the coefficients $C_k$ are found from the normalization condition, and $\mu_k$ should satisfy the equation
\begin{equation}\label{muk}
\left(\mu_k+\frac{4\delta^2}{(1-\delta^2)\mu_k}\right)\tan\mu_k=\frac{4\delta^2}{1-\delta^2}\,.
\end{equation}
Note that if $\mu_k$ is the root of Eq.~\eqref{muk}, then $-\mu_k$ will be the root of the same equation with the same function $y_k(x)$. Therefore, we can consider only the non-negative roots of Eq.~\eqref{muk}. There is infinite number of non-negative roots of Eq.~\eqref{muk}, which we will arrange in ascending order of the index $k$: $\mu_0<\mu_1<\dots$.

Further, we cannot identify the functions $y_k(x)$ with the desired functions $g_k(x)$, because the functions $y_k(x)$ are not orthogonal. We will use the following orthogonalization procedure. We take $g_0(x)=y_0(x)$, while all other $g_k(x)$ are constructed by using the following recurrent formula
\begin{equation}\label{gkreq}
g_k(x)=b_k\left[y_k(x)-\sum_{j=0}^{k-1}\left(\int\limits_0^1dx'\,g_j(x')y_k(x')\times g_j(x)\right)\right]\!,
\end{equation}
where the coefficients $b_k$ are found from the normalization condition. It is easy to see that the functions $g_k(x)$ constructed in this way are orthonormal.

Finally, we observe that the smallest non-negative root of Eq.~\eqref{muk} is $\mu_0=0$. The corresponding function $g_0(x)=y_0(x)$ is obtained by taking the limit $\mu_0\to0$ in the formula~\eqref{yk1}. As a result, we obtain
\begin{equation}\label{y0}
y_0(x)=C_0\left[\frac{2x\delta}{1-\delta}+1\right].
\end{equation}
It is easy to verify from this equation and equation~\eqref{FtoG} that $f_0(r)=\text{const}$, or, taking into account the normalization condition, $f_0(r)=\sqrt{4\pi}$.

%\bibliographystyle{apsrevlong_no_issn_url}
%\bibliography{ResonanceReferences}

\end{document}